\documentclass[aps,pra,reprint,amsmath,amssymb,superscriptaddress,onecolumn,longbibliography, notitlepage,nofootinbib]{revtex4-2}
\usepackage{mathtools}
\usepackage[hidelinks]{hyperref}
\usepackage{svg}
\usepackage{caption}
\usepackage{soul}
\usepackage{graphicx}
\usepackage{float}
\usepackage{subcaption}
\usepackage[braket, qm]{qcircuit}
\usepackage{physics}
\usepackage{xurl}
\usepackage{siunitx}
\DeclareGraphicsExtensions{.png,.pdf}

\usepackage{lipsum}
\newcommand\blfootnote[1]{%
	\begingroup
	\renewcommand\thefootnote{}\footnote{#1}%
	\addtocounter{footnote}{-1}%
	\endgroup
}

\usepackage{xcolor}

\AtBeginDocument{\RenewCommandCopy\qty\SI}

\begin{document}
	\title{Hamiltonian-reconstruction distance as a success metric\\for the Variational Quantum Eigensolver}
	
	\author{
		Leo~Joon Il~Moon$^{1,\ast,\dagger}$, 
		Mandar~M.~Sohoni$^{1,\ast,\dagger}$, 
		Michael~A.~Shimizu$^{1}$,
		Praveen~Viswanathan$^{1}$,
		Kevin~Zhang$^{2}$,
		Eun-Ah~Kim$^{2,\dagger}$
		Peter~L.~McMahon$^{1,3,\dagger}$ \\
		\textit{
			\newline
			\normalsize{$^1$School of Applied and Engineering Physics, Cornell University, Ithaca, NY, USA}\\
			\normalsize{$^2$Laboratory of Atomic and Solid State Physics, Cornell University, Ithaca, NY, USA}\\
			\normalsize{$^3$Kavli Institute at Cornell for Nanoscale Science, Cornell University, Ithaca, NY, USA}\\
	}}
	
	\begin{abstract}
		The Variational Quantum Eigensolver (VQE) is a hybrid quantum-classical algorithm for quantum simulation that can be run on near-term quantum hardware. A challenge in VQE---as well as any other heuristic algorithm for finding ground states of Hamiltonians---is to know how close the algorithm's output solution is to the true ground state, when we do not know what the true ground state or ground-state energy is. This is especially important in iterative algorithms such as VQE, where we need to decide when to stop iterating. Stopping when the current solution energy no longer changes rapidly from one iteration to the next is a common choice but in many situations can lead to stopping too early and outputting an incorrect result. Recent developments in Hamiltonian reconstruction---the inference of a Hamiltonian given an eigenstate---give a tool that can be used to assess the quality of a variational solution to a Hamiltonian-eigensolving problem: a metric quantifying the distance between the problem Hamiltonian and the reconstructed Hamiltonian can give an indication of whether the variational solution is an eigenstate, which can (as a special case) indicate if the variational solution is not the ground state. Crucially, computing the Hamiltonian-reconstruction distance does not rely on knowing the true ground state or ground-state energy. We propose and study using the Hamiltonian-reconstruction distance as a metric for assessing the success of VQE eigensolving. In numerical simulations and in demonstrations on a cloud-based trapped-ion quantum computer, we show that for examples of both  one-dimensional transverse-field-Ising (11 qubits) and two-dimensional $J_1$--$J_2$  transverse-field-Ising (6 qubits) spin problems, the Hamiltonian-reconstruction distance gives a helpful indication of whether VQE has yet found the ground state or not. Our experiments included cases where the energy plateaus as a function of the VQE iteration, which could have resulted in erroneous early stopping of the VQE algorithm, but where the Hamiltonian-reconstruction distance correctly suggests to continue iterating. We find that Hamiltonian-reconstruction distance has a useful correlation with the fidelity (i.e., state overlap) between the VQE solution and the true ground state, and with the energy difference between VQE solution's energy and the true ground-state energy. Our work suggests that Hamiltonian-reconstruction distance may be a useful tool for assessing success in VQE, including on noisy quantum processors in practice.
		
	\end{abstract}
	
	\maketitle
	\blfootnote{
		\normalsize{* These authors contributed equally.}\newline
		\normalsize{$^{\dagger}$ To whom correspondence should be addressed: \newline jm2239@cornell.edu, mms477@cornell.edu, eun-ah.kim@cornell.edu, pmcmahon@cornell.edu}
	}
	
	\vspace{-10ex}
	
	\section*{Introduction}
	\label{sec:intro}
	
	The variational quantum eigensolver (VQE) \cite{peruzzo2014variational, mcclean2016theory} is a hybrid quantum-classical algorithm that has been widely explored \cite{cerezo2021variational} in the noisy intermediate-scale quantum (NISQ) \cite{preskill2018quantum} era for quantum simulations. Its purpose is to heuristically find the ground state of a given quantum system, specified by a Hamiltonian, and to output properties of the ground state, such as its energy. The VQE combines the use of a quantum computer to prepare a (variational) trial state and a classical optimizer to adjust the trial state using a variational approach. The trial state is parameterized (from here onwards we refer to these parameters as $\bar{\theta}$) and the classical optimizer is used to find the parameters that minimize the measured value of the energy of the trial state with respect to a given target Hamiltonian. Typically, VQEs are used to find upper bounds to the ground-state energy of Hamiltonians of interest. These upper bounds are often treated as approximations to the ground-state energy, which is a central quantity of interest in both quantum chemistry \cite{deglmann2015application} and condensed-matter physics \cite{bravo2020scaling}. 
	
	A major challenge in VQE---as well as any other heuristic algorithm for finding ground states of Hamiltonians---is to know how close the algorithm's output solution is to the true ground state, when the true ground state and ground-state energy are unknown. This is especially important when one wants to avoid erroneous early termination of iterative algorithms like VQE. A first step to address this is carefully designing the skeleton of the variational quantum circuit, i.e., an ansatz \cite{tilly2022variational}. A well-designed ansatz gives the parameterized quantum circuit the ability to represent the ground state of the target Hamiltonian. Ansatz design typically considers the Haar measure, which quantifies the distribution of unitaries that can be represented by the ansatz \cite{sim2019expressibility}. However, even with a well-designed ansatz, diagnosing the performance of the VQE remains a challenge due to the lack of a well-defined, experimentally accessible metric that quantifies how close the variational wavefunction is to the target ground state. In the recent past, a few algorithms have been developed that, given an eigenstate, can reconstruct a parent Hamiltonian by measuring certain operators and the correlations between those operators \cite{qi2019determining, turkeshi2020parent, greiter2018method, bairey2019learning, zhang2022hamiltonian, chertkov2018computational}.
	
	\begin{figure}[H]
		\includegraphics[width = 1.0\textwidth]{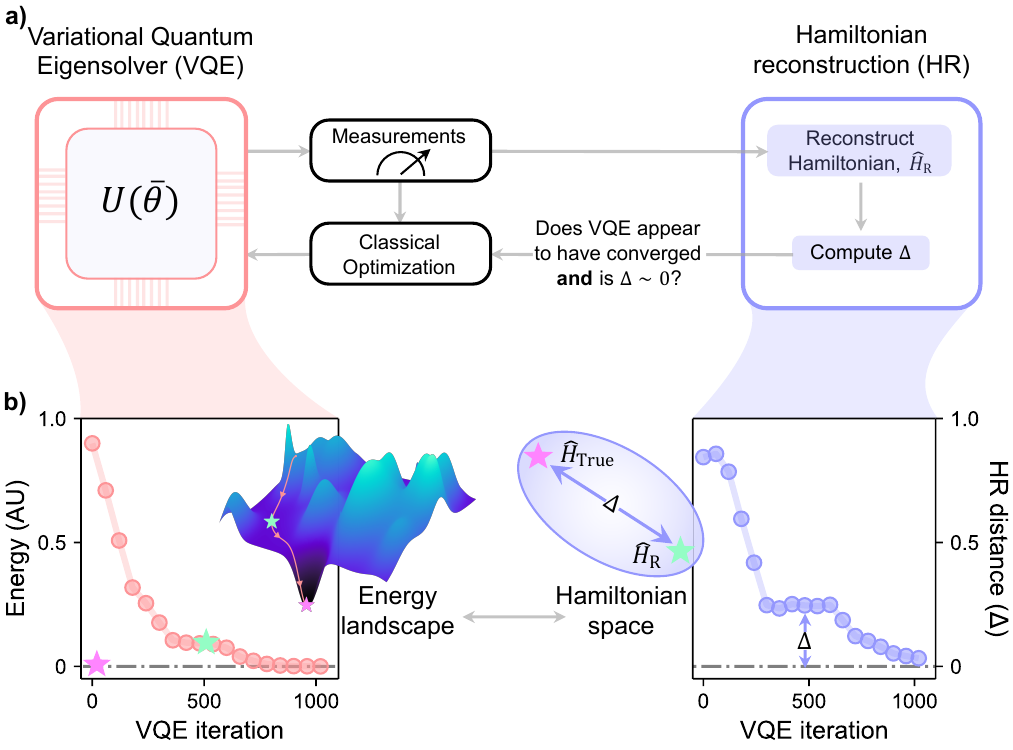}
		\caption{\textbf{A Hamiltonian reconstruction (HR) metric for assessing variational quantum eigensolver (VQE) solutions.} As VQE progresses and the variational parameters ($\bar{\theta}$) are updated, certain operators and their correlators are measured. A parent Hamiltonian, $\hat{H}_{\text{R}}(\bar{\theta})$, is reconstructed from these measurements, and the L2 distance $\Delta(\bar{\theta})$ between the reconstructed Hamiltonian, $\hat{H}_{\text{R}}(\bar{\theta})$, and the true Hamiltonian, ${\hat{H}_{\text{True}}}$, is computed to use as a metric to evaluate the performance of the VQE solver. \textbf{a)}, An overview of how the HR distance can guide VQE optimization. \textbf{b)}, A numerical simulation of an 11-qubit 1D transverse-field-Ising model (1D-TFIM) VQE optimization where the HR distance suggests that the VQE solution isn't optimal even though it appears to have converged (green star). More details about this example numerical simulation can be found in Appendix A.}
		\label{Fig1}
	\end{figure}
	
	In this work, we build on these previously proposed theoretical Hamiltonian reconstruction algorithms to define a metric called the Hamiltonian Reconstruction (HR) distance, $\Delta(\bar{\theta})$ (see Figure \ref{Fig1}a) to assess VQE solutions. Figure \ref{Fig1}b demonstrates a toy simulation where the HR distance suggests that the VQE solution isn't optimal even though it appears to have converged. More details about this scenario can be found in Appendix A and Appendix Figure 1. To quantitatively demonstrate that the HR distance has the potential to be an effective metric for diagnosing VQE performance, we look at the correlation between the experimentally measured HR distance and simulated fidelity to the ground state (henceforth referred to as fidelity) for an 11-qubit 1D-transverse-field Ising model (TFIM) and a 6-qubit $J_{1}$-$J_{2}$ TFIM (2$\times$3 lattice) on a system of trapped ions (IonQ Harmony \cite{wright2019benchmarking}). By measuring only a polynomially expensive number of operators (in terms of the number of operators in the target Hamiltonian), we compute HR distances for both models. To understand how the HR distance is affected by 1- and 2-qubit gate error probabilities, and its relation to fidelity, we run extensive simulations to look at the correlation between the HR distance and fidelity of 1D-TFIM models in different noisy conditions. 
	
	Our results suggest that the HR distance has a positive correlation with the energy and a negative correlation with the fidelity when: (1) the VQE optimization is close to convergence; (2) the energy gap between the ground state and the first few eigenstates is sufficiently large (typically large enough so that the variance in energy measurements due to gate infidelity, readout noise and shot noise don't hinder VQE convergence); and (3) the gate fidelities are sufficiently high. Given the multitude of approaches in designing variational algorithms, in particular, ansatz design with machine learning \cite{ostaszewski2021reinforcement, kuo2021quantum}, we expect that the HR distance could aid in designing cost functions and ansatzes for variational quantum circuits. 
	
	\section*{The Hamiltonian reconstruction distance}
	
	Figure \ref{Fig1}a shows an overview of how the HR distance, $\Delta$, is measured during VQE optimization. In this work, we consider an 11-qubit 1D-TFIM Hamiltonian and a 6-qubit $J_{1}$-$J_{2}$ TFIM (only with Pauli-Z couplings for the nearest and the next-nearest neighbors) on a $3\times 2$ lattice (see spin diagrams in Figures \ref{Fig2}a and \ref{Fig2}c for reference). The Hamiltonians are given as follows:
	
	\begin{gather}
		\hat{H}_{\text{1D-TFIM}} = \sum_{i=1}^{n} \sigma_i^{x} + J\sum_{i=1}^{n-1} \sigma_{i}^{z}\sigma_{i+1}^{z} \label{eq:1} \\
		\hat{H}_{J_{1}-J_{2}\text{TFIM}} = \sum_{i=1}^{n} \sigma_i^{x} + J_{1}\sum_{\ev{i, j}} \sigma_{i}^{z}\sigma_{j}^{z} + J_{2}\sum_{\ev{\ev{i, j}}} \sigma_{i}^{z}\sigma_{j}^{z} \label{eq:2}
	\end{gather}
	
	where $\sigma_{i}^{x}$ and $\sigma_{i}^{z}$ correspond to the Pauli-X and Pauli-Z operators on the $i^{th}$ spin in the chain/lattice, $J$ is the nearest-neighbor spins' coupling strength for the 1D-TFIM, and $\ev{i, j}$ and $\ev{\ev{i, j}}$ denote nearest and next-nearest neighbors with $J_{1}$ and $J_2$ being their corresponding coupling strengths. For all experiments and simulations, we used $J = 0.5$, $J_{1} = 0.5$ and $J_{2} = 0.2$. The values of $J_{1}$ and $J_{2}$ were chosen to be close to the maximally frustrated point for the spin-$\frac{1}{2}$ $J_{1}$-$J_{2}$ Heisenberg model \cite{choo2019two, dagotto1989phase}.
	
	To perform Hamiltonian reconstruction \cite{zhang2022hamiltonian}, one must first decide on a set of operators, $\{\hat{H}_{i}\}$, whose span contains the true Hamiltonian. The Hamiltonian reconstruction protocol, described as follows, then returns a linear combination of these chosen operators, $\hat{H}_{\text{R}} = \sum_{i}\tilde{c}_{i}\hat{H}_{i}$. During VQE optimization, a covariance matrix, $Q(\bar{\theta})$, of these operators is constructed by measuring them, $\ev{\hat{H}_{i}}$, and their correlators, $\ev{\hat{H}_{i}\hat{H}_{j}}$. The eigenvector, $\{\tilde{c}_{i}(\bar{\theta})\}$, corresponding to the lowest eigenvalue of this covariance matrix results in the reconstructed Hamiltonian. The HR distance, $\Delta(\bar{\theta}) = \lVert \bar{\tilde{c}}_{i}(\bar{\theta}) - \bar{c}\rVert_{2}$, is found by computing the euclidean distance between the coefficients of the reconstructed Hamiltonian and the coefficients of the true Hamiltonian, $\hat{H}_{\text{True}} = \sum_{i}c_{i}\hat{H}_{i}$. A detailed description of the Hamiltonian reconstruction process can be found in Appendix B.
	
	In our simulations and experiments, when finding the HR distance for the 1D-TFIM and the $J_{1}$-$J_{2}$ TFIM, we choose the set of operators, $\{ \hat{H}_{i} \}$, to be $\{ \sum_{i=1}^{n} \sigma_i^{x}, \quad \sum_{i=1}^{n-1} \sigma_{i}^{z}\sigma_{i+1}^{z}\}$ and $\{ \sum_{i=1}^{n} \sigma_i^{x}, \quad \sum_{\ev{i, j}} \sigma_{i}^{z}\sigma_{j}^{z}, \quad \sum_{\ev{\ev{i, j}}} \sigma_{i}^{z}\sigma_{j}^{z}\}$ respectively. This choice allows us to search for different nearest neighbor and next-nearest neighbor coupling strengths. More details about why these operators were chosen can be found in Appendix C (also see Appendix Figure 2).
	
	\section*{Results}
	Both the 11-qubit 1D-TFIM and the 6-qubit $J_{1}$-$J_{2}$ TFIM cloud-based experiments were run on a system of trapped ions (IonQ's quantum backend, Harmony \cite{wright2019benchmarking}). An alternating-layered-ansatz (ALA) \cite{cerezo2021cost} was used in both cases. We chose the ALA for our experiments because it uses fewer CNOT gates which reduces state preparation errors. The ALA used for the 1D-TFIM had 33 trainable parameters, each being the phase of an RY gate, and 15 CNOT gates (see Appendix Figure 3). The ALA used for the $J_{1}$-$J_{2}$ TFIM consisted of 18 trainable parameters, each being the phase of an RY gate, and 8 CNOT gates (see Appendix Figure 4). We'd like to mention here that the ALA is not an optimal ansatz for finding the ground state of these hamiltonians, and a much better ansatz exists (see Appendix D and Appendix Figure 5). Before collecting the experimental data, we first, in simulation, ran a VQE protocol for both the Hamiltonians (see Appendices E and F). The VQE for the 11-qubit 1D-TFIM contained about 1400 iterations, while the VQE for the 6-qubit $J_{1}$-$J_{2}$ TFIM had about 500 iterations. 
	
	\begin{figure}[H]
		\centering
		\includegraphics[width = \textwidth]{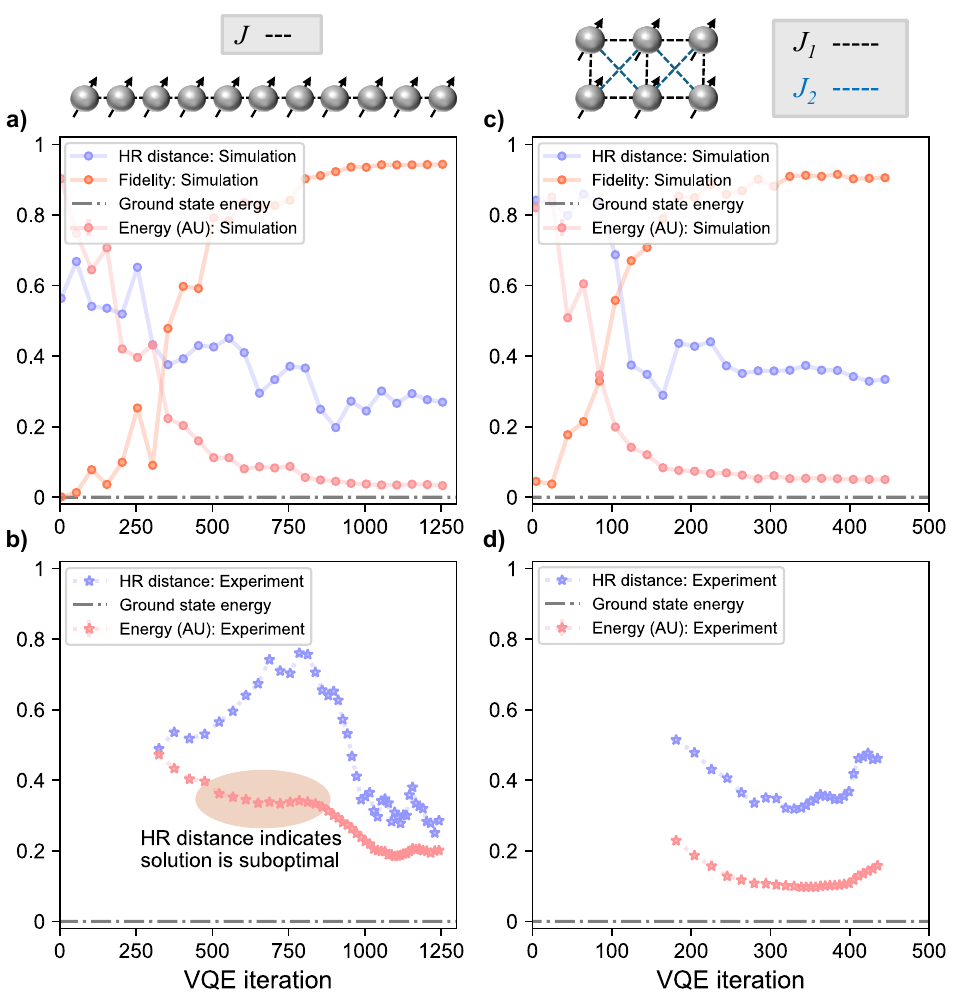}
		\caption{\textbf{Performance of the Hamiltonian reconstruction (HR) distance metric for an 11-qubit 1D-TFIM and a 6-qubit $J_{1}$-$J_{2}$ TFIM (experiments on a cloud-based quantum computer and numerical simulations, as labeled).} \textbf{a)}, Simulated HR distance (purple circles), simulated energy (pink circles), and simulated fidelity (orange circles) as a function of the VQE progression for the 11-qubit 1D-TFIM. This simulation was run using an ALA (see Appendix Figure 3), 4000 shots, and no depolarization noise. \textbf{b)}, Experimentally measured HR distance (purple stars), experimentally measured energy (pink stars) for certain iterations from \textbf{a}. \textbf{c)}, Simulated HR distance (purple circles), simulated energy (pink circles), and simulated fidelity (orange circles) as a function of the VQE progression for the 6-qubit $J_{1}$-$J_{2}$ TFIM. This simulation was run using an ALA (see Appendix Figure 4), 10000 shots, and no depolarization noise. \textbf{b)}, Experimentally measured HR distance (purple stars), experimentally measured energy (pink stars) for certain iterations from \textbf{c)}. For both \textbf{b} and \textbf{d}, VQE was first simulated, and then using the parameter list, $\{\bar{\theta}\}$, obtained, experiments were run for certain iterations. The coupling value used for the 1D-TFIM was $J = 0.5$, the values used for the $J_{1}$-$J_{2}$ TFIM were $J_{1} = 0.5$ and $J_{2} = 0.2$. All experiments were performed on IonQ's quantum backend, Harmony \cite{wright2019benchmarking}. A 14-point moving average was used for the experimental data to better show trends. More details about how the experiments and simulations were run and the raw data can be found in the results section and Appendix E respectively.}
		\label{Fig2}
	\end{figure}
	
	\noindent
	We then chose parameters, $\bar{\theta}$, corresponding to 58 points (40 points) in the simulated VQE to run on actual quantum hardware for the 11-qubit 1D-TFIM (6-qubit $J_{1}$-$J_{2}$ TFIM). The VQE simulations were run with energy as the cost function and with shot-noise only. 4000 shots were used per experimental HR measurement for the 11-qubit 1D-TFIM and 10000 shots were used per experimental HR measurement for the 6-qubit $J_{1}$-$J_{2}$ TFIM.
	
	Figure \ref{Fig2}a shows the energy, HR distance, and fidelity in simulation as a function of the VQE iterations for the 11-qubit 1D-TFIM with no depolarization noise (we used 10000 shots for the energy and 4000 shots for the HR distance). Notice that, while the VQE appears to have converged between iterations 600 and 700, the HR distance fluctuates around a value of 0.4, thus indicating that the prepared state is not optimal. Even around iteration 1250, when the energy stopped decreasing and the optimizer had reached its termination condition, the HR distance's non-zero value indicates that the state preparation is sub-optimal (reflected by the fidelity only being $\sim$ 90\%). Figure \ref{Fig2}b shows the experimentally measured energy (pink stars) and experimentally measured HR distance (purple stars) for certain chosen VQE iterations from \ref{Fig2}a (as mentioned earlier). Again, one can see that the HR distance indicates that the quality of the VQE solution between iterations 500 and 750 is much worse than near iteration 1250, even though the VQE appears to have converged around iteration 700. We note that a 14-point moving average was used for the experimental data to better show the energy and HR distance trends. The raw data can be found in Appendix Figure 6. Similarly, Figure \ref{Fig2}c shows the energy, HR distance, and fidelity in simulation as a function of the VQE iterations for the 6-qubit $J_{1}$-$J_{2}$ TFIM with no depolarization noise and measurements with 10000 shots. Figure \ref{Fig2}d shows the experimentally measured energy (pink stars) and experimentally measured HR distance (purple stars) for certain chosen VQE iterations from \ref{Fig2}c. As with \ref{Fig2}b, a 14-point moving average was used for the experimental data. The raw data can be found in Appendix Figure 7.
	
	To understand how reliable the HR distance's assessment of the VQE solution quality is, especially with NISQ hardware, we ran simulations to see the effect of 1- and 2-qubit gate error probabilities on the HR distance, and the correlation between the HR distance and fidelity. For these simulations we used an ansatz that was better at finding the ground state of the 11-qubit 1D-TFIM than the ALA used in experiment (see Appendix Figure 5 for more details). Figures \ref{Fig3}a and \ref{Fig3}b show the effect, in simulation, of 1- and 2-qubit gate error probabilities on the HR distance and fidelity for the 11-qubit 1D-TFIM. We first simulated a VQE and then for each pair of 1- and 2-qubit gate infidelities we computed the expected value of the HR distance (i.e., the infinite shot limit) and fidelity at convergence, i.e., when the optimizer attained termination conditions. Figure \ref{Fig3}c shows the energy, HR distance, and fidelity of the simulated VQE for three different 1- and 2-qubit gate error probability values (subplots (i), (ii) and (iii)). In Figure \ref{Fig3}c, subplots (i), (ii) and (iii), the left panels show the VQE trajectories, and the right panels show a scatter plot between the fidelity and HR distance. The 1- and 2-qubit gate error probabilities used in \ref{Fig3}c (i), (ii), and (iii) are ($\num{1.08}\times 10^{-6}$, $\num{1.05}\times 10^{-4}$), ($\num{2.86}\times 10^{-3}$, $\num{1.91}\times 10^{-2}$), and ($\num{2.69}\times 10^{-2}$, $\num{8.40}\times 10^{-2}$) respectively. With very low gate error probabilities, and for this simple Hamiltonian, there seems to be a strong correlation between the HR distance and fidelity. However, as the errors start to increase, the correlation quickly vanishes. We'd like to point out here that care must be taken when comparing the HR distance across states with different effective depolarization noises because we've empirically found that the correlation between the fidelity and HR distance changes as the depolarization noise increases. For instance an HR distance value of 0.25 means different things for a circuit with low depolarization noise as compared to one with high depolarization noise. Further analysis about how depolarization noise affects the correlation between the HR distance and the fidelity can be found in Appendix G and Appendix Figures 8 and 9.
	
	Finally, to see how shot noise affects the HR distance, we ran simulations for both the 11-qubit 1D-TFIM and the 6-qubit $J_{1}$-$J_{2}$ TFIM. Appendix Figure 10 shows, in simulation, the standard deviation of the HR distance as a function of the number of shots used for measurement. The ansatz used for the simulation was the ALA (the same as the one used for the experiments in Figures \ref{Fig2}b and \ref{Fig2}d). One can see that, for these models, a large number of shots ($\sim 10^{4}$) is required to have a reasonable signal-to-noise ratio (SNR) for the HR distance. Appendix H goes into detail about the perturbation-robustness of the HR distance.
	
	While all of these results suggest that the HR distance has the potential to diagnose the quality of VQE solutions, there are certain conditions under which the HR distance can be reliably interpreted. The first of these is that when using energy as the cost function, the VQE algorithm must be close to convergence before the HR distance is a reliable indicator of solution quality. The reason for this is because the HR distance will be zero when the parameterized wavefunction is \textit{any} eigenstate of the target Hamiltonian. Thus, early on in the optimization, a low HR distance might not be indicative of closeness to the ground state (see iterations 1-200 in Appendix Figures 11a and 11b). Another condition that we've found when simulating noisy hardware is the HR distance's sensitivity to the energy gap between the ground state and the first few excited states. Appendix I goes into detail about the effect of the gap between the first excited state and the ground state, and the HR distance. We empirically find that when this gap is small, the HR distance shows negative correlations with the fidelity to higher excited states in addition with fidelity to the ground state. This indicates that noisy hardware and a small energy gap might hinder diagnosis.
	
	\section*{Discussion}
	We defined a metric, termed the Hamiltonian reconstruction (HR) distance, and demonstrated that it can be useful in assessing the success of Variational Quantum Eigensolver (VQE) algorithms on noisy intermediate-scale quantum (NISQ) hardware without any knowledge of the true ground state or ground-state energy.
	
	\begin{figure}[H]
		\includegraphics[width = 1.0\textwidth]{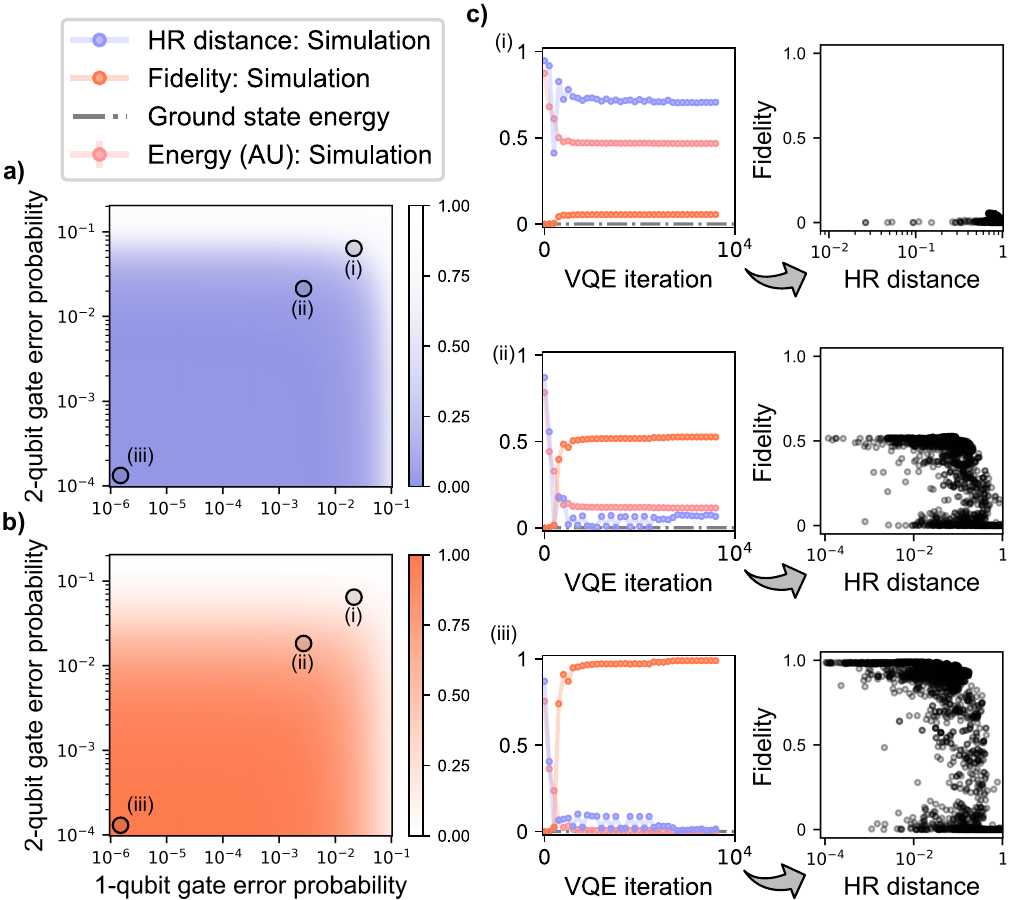}
		\caption{\textbf{Noise dependence of the HR distance metric and correlation with fidelity (numerical simulations).} \textbf{a)}, Effect of 1- and 2-qubit gate error probabilities on HR distance for the 11-qubit 1D-TFIM VQE. \textbf{b)}, Effect of 1- and 2-qubit gate error probabilities on fidelity for the 11-qubit 1D-TFIM VQE. \textbf{c)}, Correlation between the fidelity and HR distance for three different 1- and 2-qubit gate error probability values (the panels on the right in i, ii, and iii) as shown in \textbf{a)} and \textbf{b)}. The panels on the left in \textbf{c)} (i), (ii), and (iii) show the energy (pink cirlces), HR distance (purple circles) and fidelity (orange circles) as VQE progeresses. The panels on the right in \textbf{c)} (i), (ii), and (iii) show a scatter plot between the fidelity and HR distance. The 1- and 2-qubit gate error probability values used in \textbf{c)} (i), (ii), and (iii) are ($\num{1.08}\times 10^{-6}$, $\num{1.05}\times 10^{-4}$), ($\num{2.86}\times 10^{-3}$, $\num{1.91}\times 10^{-2}$), and ($\num{2.69}\times 10^{-2}$, $\num{8.40}\times 10^{-2}$) resplectively.}
		\label{Fig3}
	\end{figure}
	
	Our investigation included both quantum-hardware experiments and circuit simulations, focusing on two distinct Hamiltonians: an 11-qubit 1D-transverse-field Ising model (TFIM) and a 6-qubit $J_{1}$-$J_{2}$ TFIM (lattice dimensions 3x2). Our findings suggest that the HR distance tends to converge to lower values as the VQE optimization process progresses, and that it has a negative correlation with the fidelity to the ground state (close to VQE convergence). This suggests that the HR distance can serve as a reliable indicator of optimization progress when the true ground-state energy is unknown, potentially assisting in the determination of when the VQE algorithm has reached a satisfactory solution.
	
	Looking to the future, the HR distance could be useful as a metric to include in multi-objective cost functions for optimization problems on NISQ hardware. Typically the sole objective used in VQE is to minimize the energy. However, including the HR distance as an additional objective in the cost function could potentially improve the convergence speed. The quality of the solutions found if the true ground state is not reached could also improve: by minimizing the HR distance alongside energy, VQE may find states that more closely resemble the ground state, even if they have the same energy, thus enhancing the fidelity of the solution. Given the recent developments in quantum multi-objective optimization \cite{abbas2023quantum, boyd2022training} and the success that multi-objective cost functions have had in physics-informed machine learning \cite{bischof2021multi}, we believe that this approach could lead to improved VQE performance \cite{li2022efficient}, especially in situations where the ground state is challenging to reach due to barren plateaus \cite{mcclean2018barren} in a single objective.
	
	The variance of the Hamiltonian has been suggested \cite{zhang2022variational} as an alternative metric to use in the cost function for VQE optimization (see Appendix J and Appendix Figure 12). While this has been effective, a major drawback with the variance is that comparison between different variational wavefunctions (for example, comparing different ans\"atze for a single Hamiltonian) becomes a case-by-case exercise (see Appendix Figured 6 and 7 for a comparison between the HR distance and the Hamiltonian variance for the 11-qubit 1D-TFIM and the 6-qubit $J_{1}$-$J_{2}$ TFIM). We believe the HR distance provides two additional benefits, beyond those we have already discussed---(1) control of the reconstruction subspace (by choosing the operators used in construction) which, in principle, could allow for tailored cost functions; and (2) a way to compare different output states when VQE is close to convergence (for the same Hamiltonian) because it obeys the properties of a metric.
	
	Arguably the most compelling features of the HR distance are that it is efficient to compute (requiring measuring only polynomially many operators in the number of terms in the Hamiltonian), and that as a success metric for VQE it does not require prior knowledge of the ground-state energy or other properties of the ground state. This makes it a practical choice for use with VQE in the way we have demonstrated in this paper, but there is also the potential to incorporate the HR distance into variants of VQE in new ways, such as in multi-objective optimization.
	
	\section*{Data and code availability}
	All data generated and code used in this work are available at: \url{https://zenodo.org/records/10822811}. We also provide a pedagogical implementation of our method at \url{https://github.com/mcmahon-lab/Hamiltonian-Reconstruction-Metric}, which may be of use to anyone wanting to apply the technique we describe to their own problems.
	
	\section*{Acknowledgements}
	The authors wish to thank thank Eliott Rosenberg, Vladimir Kremenetski, Abhishek Kejriwal, and Martin Stein for helpful discussions.  Portions of this work were supported by the National Science Foundation (Award No. CHE-2038027 to E.A.K. and P.L.M.). P.L.M. acknowledges membership of the CIFAR Quantum Information Science Program as an Azrieli Global Scholar. 
	K.Z. acknowledges support by the NSF under EAGER OSP-136036 and NSERC under a PGS-D scholarship.
	E.A.K. acknowledges support by the NSF under OAC-2118310, EAGER OSP-136036, the Ewha Frontier 10-10 Research Grant, and the Simons Fellowship in Theoretical Physics award 920665, the Gordon and Betty Moore Foundation’s EPiQS Initiative, Grant GBMF10436, and a New Frontier Grant from Cornell University’s College of Arts and Sciences. 
	We gratefully acknowledge access to IonQ's 11-qubit device -- Harmony, facilitated by Fabrice Frachon. The views expressed are those of the authors, and do not reflect the official policy or position of IonQ.
	
	\section*{Author Contributions}
	L.M., M.A.S, P.V. and M.M.S. ran the simulations. L.M. collected the data. M.M.S. and L.M. analyzed the data. M.M.S., L.M., and K.Z. designed the Hamiltonian reconstruction metric. P.L.M. and E.A.K. conceived and supervised the project. All authors contributed to writing the manuscript.
	
	\bibliographystyle{mcmahonlab}
	\bibliography{references_main}

\end{document}


\setcounter{page}{1}
	\title{Appendix for Hamiltonian-Reconstruction distance as a success metric for the Variational Quantum Eigensolver}
	\maketitle
	
	\tableofcontents
	\clearpage
	
	\section{An example of how HR distance can diagnose the quality of a VQE solution}
	
	\begin{figure}[H]
		\includegraphics[width = 1.0\textwidth]{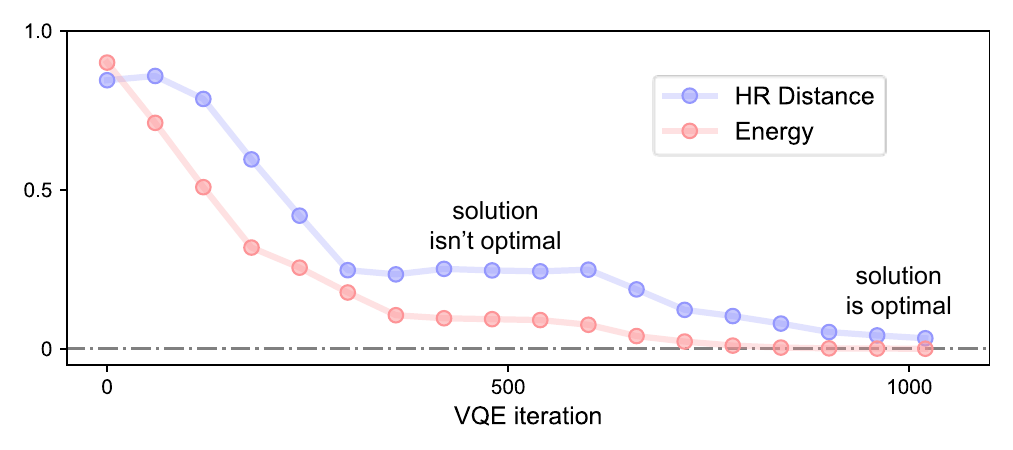}
		\caption{An example of a scenario, in simulation, of an 11-qubit 1D transverse-field-Ising model VQE optimization, where the HR distance suggests that the VQE solution isn't optimal even though it appears to have converged.}
		\label{supp_fig_1}
	\end{figure}
	
	We demonstrate a scenario where measuring the HR distance does help in diagnosing a VQE solution. We simulated VQE for an 11-qubit 1D-TFIM Hamiltonian with $J = 0.5$ (\ref{eq:4}) without shot noise and depolarization noise. We used an 11-qubit 2 layer YY ansatz (Appendix Figure \ref{supp_fig_5}b). Appendix Figure \ref{supp_fig_1} shows a regime (between iterations 375 and 650) where the VQE appears to have converged to either a local minimum or appears to be stuck in a barren plateau. If one was only measuring the energy and did not know the true ground state energy, it's possible that the VQE might have been terminated when the solution was sub-optimal. However, the HR distance in this regime had a value reasonably far from zero ($\approx 0.25$), thus indicating that the solution was sub-optimal. When the VQE truly converges, one can see that the HR distance goes closer to 0. This indicates that the HR distance can be used as a diagnostic for VQE by providing additional information that the energy does not provide.
	
	\section{The Hamiltonian Reconstruction Distance}
	
	We use Hamiltonian Reconstruction (HR) \cite{qi2019determining, chertkov2018computational} to measure how well VQE optimization approximated the ground state of the Hamiltonian. To do so, we first obtain the reconstructed Hamiltonian by making measurements on the output state $\rho$ from VQE optimization and then compare the reconstructed Hamiltonian with the true Hamiltonian using the L2 distance. (\ref{eq:1}) illustrates steps to obtain the reconstructed Hamiltonian.
	
	\begin{equation}
		\begin{aligned}
			\hat{H_{R}}[\{ \tilde{c}_{i}\}] = \sum_{i}^{N}\tilde{c}_{i}\hat{H_{i}}, \text{where } \hat{H_{i}} \text{ is the $i^{th}$ operator}\\
			Q[\rho;\hat{H}]_{ij} = \frac{1}{2}(\langle \hat{H_{i}}\hat{H_{j}} \rangle + \langle \hat{H_{j}}\hat{H_{i}} \rangle) - \langle \hat{H_{i}}\rangle \langle \hat{H_{j}}\rangle \\ 
			Q[\rho; \hat{H}]\text{ is the covariance matrix}\\
			Q[\rho; \hat{H}] = V^{T}DV, V = \begin{bmatrix}\vec{v_1} & \vec{v_2} &...& \vec{v_{N}}\end{bmatrix}, D = diag(\sigma_{1}, \sigma_2 ... \sigma_N)\\
			\text{Entries of }\vec{v_{1}} \text{ corresponds to coefficients } \{ \tilde{c}_{i} \} \label{eq:1}
		\end{aligned}
	\end{equation}
	
	Here, $\hat{H_{R}}[\{\tilde{c}_{i}\}]$ is the reconstructed Hamiltonian. $N$ is the number of operators used in the reconstruction. $\vec{v_{i}}$ is the $i^{th}$ eigenvector of the covariance matrix $Q[\rho; \hat{H}]$. $\sigma_{i}$ is its corresponding eigenvalues, where $\sigma_{1} \leq \sigma_{2} ... \leq \sigma_{N}$. $D = diag(\sigma_{1}, \sigma_2 ... \sigma_N)$ is a diagonal matrix with diagonal values $\sigma_{1}, \sigma_{2} ... , \sigma_{N}$. Lastly, $\langle O \rangle$ indicates the expectation value of some observable $O$.
	
	To successfully run Hamiltonian reconstruction, the true Hamiltonian must lie on the span of the chosen set of operators $\{\hat{H_{i}} \}$. Then, if $\rho$ is the exact ground state of the original true Hamiltonian $\hat{H}_{True}$, $\hat{H}_{True}$ lies in the nullspace of $Q[\rho; \hat{H}]$. Thus, if the covariance matrix of the ground state does not have degeneracy, $\hat{H}_{True}$ is equal to $\hat{H_{R}}$, as the coefficients of $\hat{H}_{True}$ is the eigenvector $\vec{v_{1}}$ of the covariance matrix $Q[\rho; \hat{H}]$ with the lowest eigenvalue \cite{zhang2022hamiltonian}. 
	
	(\ref{eq:2}) shows how we define the HR distance as the L2 distance between the reconstructed Hamiltonian and the true Hamiltonian.
	
	\begin{equation}
		\begin{aligned}
			\hat{H}_{True}[\{ c_{i}\}] = \sum_{i}^{N}c_{i}\hat{H}_{i}, \hat{H}_{R}[\{ \tilde{c}_{i}\}] = \sum_{i}^{N}\tilde{c}_{i}\hat{H}_{i} \\
			\text{(HR distance)} = \sqrt{||\begin{bmatrix} c_{1} \\ c_{2} \\ ... \\c_{N} \end{bmatrix} - \begin{bmatrix} \tilde{c}_{1} \\ \tilde{c}_{2} \\ ... \\\tilde{c}_{N} \end{bmatrix}||^{2} } = \sqrt{\sum_{i=1}^{N} (c_{i} - \tilde{c}_{i})^{2}} \\\label{eq:2}
		\end{aligned}
	\end{equation}
	
	Here, $\hat{H}_{True}[\{ c_{i}\}]$ is the true (target) Hamiltonian, and $\hat{H}_{R}[\{ \tilde{c}_{i}\}]$ is the reconstructed Hamiltonian. We normalize the coefficient vector $\{ c_{i} \}$ for the true Hamiltonian $\hat{H}_{True}$, as the diagonalization listed in (\ref{eq:1}) already normalizes the coefficient vector $\{ \tilde{c}_{i}\}$ for the reconstructed Hamiltonian $\hat{H}_{R}$. If VQE optimization outputs wavefunction $\rho$ that is equal to the actual ground state, the corresponding HR distance is equal to zero, as $\hat{H}_{true} = \hat{H}_{R}$. Furthermore, the maximum value of HR distance is one, because the coefficient vectors for both Hamiltonians are normalized.

	\section{Choosing basis operators used in Hamiltonian reconstruction}\label{section:B}
	
	One challenge in HR is choosing a set of operators to run it. There are infinite choices of set of operators that can be used, as long as the true Hamiltonian lies on the span of the set of operators we choose. 
	
	To simplify our problem of choosing the best set of operators,  we first restricted ourselves to six operators written in (\ref{eq:3}) and considered the 8-qubit 1D-transverse-field Ising model (TFIM) Hamiltonian with coupling strength $J = 0.5$.
	
	\begin{equation}
		\begin{aligned}
			\sum_{i=1}^{N-1} \sigma_i^{x}\sigma_{i+1}^{x}, \sum_{i=1}^{N-1} \sigma_i^{y}\sigma_{i+1}^{y}, \sum_{i=1}^{N-1} \sigma_i^{z}\sigma_{i+1}^{z}, \sum_{i=1}^{N} \sigma_i^{x}, \sum_{i=1}^{N} \sigma_i^{y}, \sum_{i=1}^{N} \sigma_i^{z}\\
			\label{eq:3}
		\end{aligned}
	\end{equation}
	
	We then created 100 different wavefunctions by randomly perturbing the ground state wavefunction to measure the correlation between the HR distance and the fidelity, while varying the choice of set of operators. To create 100 different wavefunctions, we first randomly chose entries of the ground state wavefunction in the computational basis and scaled them, with random values between 0 and 1. We then normalize the scaled wavefunctions and only use the wavefunctions that have fidelity squared value higher than 0.8, because our focus is to evaluate how HR distance performs as a metric when ansatzes’ parameters in VQEs are near convergence. Here, we measure the expectation values needed in Hamiltonian reconstruction by calculating $\bra{\Psi'} \hat{H}_{i}\hat{H}_{j} \ket{\Psi'}$ explicitly, where $\hat{H}_{i}$ and $\hat{H}_{j}$ are the $i^{th}$ and $j^{th}$ operators in the set of operators we choose and $\ket{\Psi'}$ is the perturbed ground state wavefunction.
	
	\begin{table}
		\begin{center}
			\begin{tabular}{ | c | c | c | } 
				\hline
				Set of Operators $\{ H_{i}\}$ & Correlation between the HR distance and the fidelity \\ 
				\hline
				\textbf{\{X, ZZ\}} & \textbf{-0.985} \\
				\hline
				\{X, ZZ, Z\} & -0.930  \\
				\hline
				\textbf{\{X, ZZ, Y\}} & \textbf{-0.985} \\
				\hline
				\{X, ZZ, XX\} & -0.395 \\
				\hline
				\{X, ZZ, YY\} & -0.693 \\
				\hline
				\{X, ZZ, Z, Y\} & -0.930 \\
				\hline
				\{X, ZZ, Z, XX\} & -0.396 \\
				\hline
				\{X, ZZ, Z, YY\} & -0.693 \\
				\hline
				\{X, ZZ, Y, XX\} & -0.395 \\
				\hline
				\{X, ZZ, Y, YY\} & -0.693 \\
				\hline
				\{X, ZZ, XX, YY\} & -0.502 \\
				\hline
				\{X, ZZ, Z, Y, XX\} & -0.396 \\
				\hline
				\{X, ZZ, Z, Y, YY\} & -0.693 \\
				\hline
				\{X, ZZ, Z, XX, YY\} & -0.504 \\
				\hline
				\{X, ZZ, Y, XX, YY\} & -0.502 \\
				\hline
				\{X, Y, Z, XX, YY, ZZ\} & -0.504\\ 
				\hline
			\end{tabular}
		\end{center}
		\caption{ Correlation between the HR distance and the fidelity from 100 randomly perturbed ground state wavefunction. Plots corresponding to each entry are in Appendix Figure \ref{supp_fig_2}. $\{\hat{H}_{i}\}$ are set of operators used in reconstruction. X, Y, Z, XX, YY, and ZZ indicate operators $\sum_{i=1}^{N} \sigma_i^{x}, \sum_{i=1}^{N} \sigma_i^{y}, \sum_{i=1}^{N} \sigma_i^{z}$, $\sum_{i=1}^{N-1} \sigma_i^{x}\sigma_{i+1}^{x}, \sum_{i=1}^{N-1} \sigma_i^{y}\sigma_{i+1}^{y}, \text{ and } \sum_{i=1}^{N-1} \sigma_i^{z}\sigma_{i+1}^{z}$ respectively.
		}
		\label{sup_tab_1}
	\end{table}
	
	Appendix Table \ref{sup_tab_1} shows how the correlation between the HR distance and the fidelity varies with respect to the choice of set of operators. We obtain the best negative correlation between the HR distance and the fidelity when we only use the operators in the Hamiltonian $\{\sum_{i=1}^{N} \sigma_i^{x}, \sum_{i=1}^{N-1} \sigma_i^{z}\sigma_{i+1}^{z}\}$ or when we use $\{ \sum_{i=1}^{N} \sigma_i^{x}, \sum_{i=1}^{N} \sigma_i^{y}, \sum_{i=1}^{N-1} \sigma_i^{z}\sigma_{i+1}^{z} \}$ to run Hamiltonian reconstruction. However, using the operators only in the Hamiltonian has an advantage over having additional operators, as we can obtain HR distances fewer number of measurements, which reduces the impact of noise in NISQ hardware and decreases the runtime of the Hamiltonian reconstruction algorithm. Thus, considering how well the HR distance performed as a metric when we only used the operators in the Hamiltonian for reconstructing the 8-qubit 1D-TFIM Hamiltonian, we also only used the operators in the true Hamiltonian to reconstruct the 11-qubit 1D-TFIM Hamiltonian and the 6-qubit $J_{1}$-$J_{2}$ TFIM Hamiltonian (on a 3 $\times$ 2 lattice) to measure HR distances.
	
	\begin{figure}
		\includegraphics[width = 1.0\textwidth]{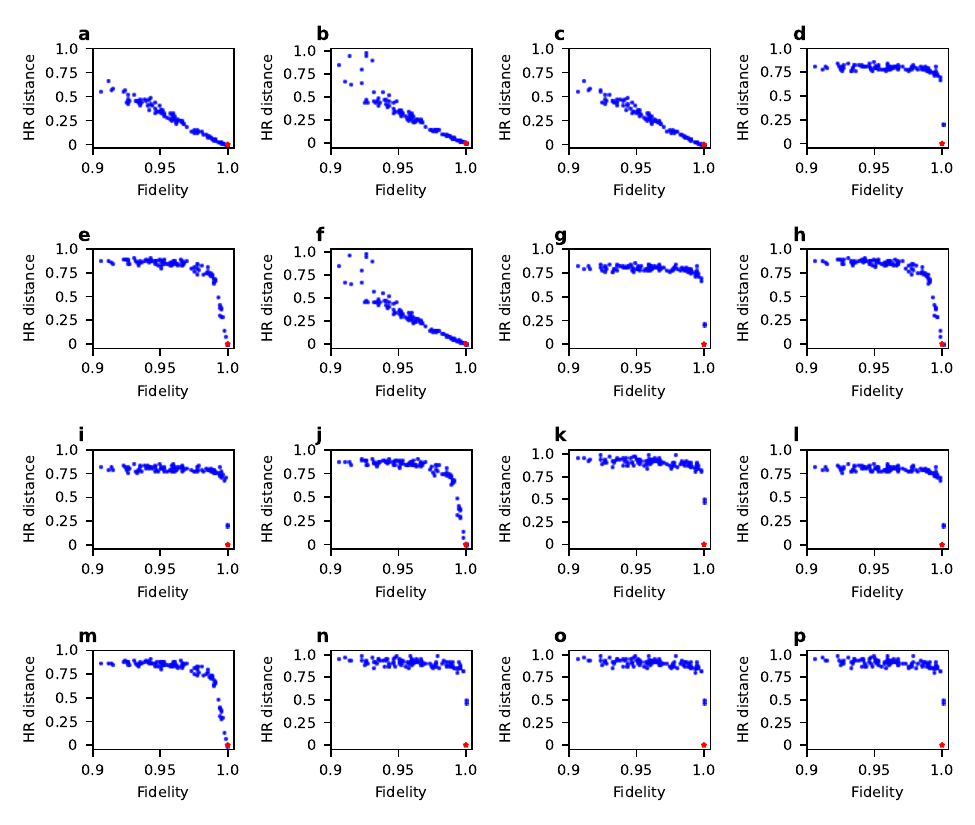}
		\caption{ 
			HR distances and fidelities of 100 randomly perturbed ground state wavefunctions for different sets of operators for Hamiltonian reconstruction. The red star in each plot indicates the fidelity and the HR distance of the ground state. Each subplot uses the following set of operators for Hamiltonian reconstruction: \textbf{a}. $\{X, ZZ\}$, \textbf{b}. $\{X, ZZ, Z\}$, \textbf{c}. $\{X, ZZ, Y\}$, \textbf{d}. $\{X, ZZ, XX\}$, \textbf{e}. $\{X, ZZ, YY\}$, \textbf{f}. $\{X, ZZ, Z, Y\}$, \textbf{g}. $\{X, ZZ, Z, XX\}$, \textbf{h}. $\{ X, ZZ, Z, YY\}$, \textbf{i}. $\{ X, ZZ, Y, XX\}$, \textbf{j}. $\{X, ZZ, Y, YY\}$, \textbf{k}. $\{X, ZZ, XX, YY\}$, \textbf{l}. $\{X, ZZ, Z, Y, XX\}$, \textbf{m}. $\{X, ZZ, Z, Y, YY\}$, \textbf{n}. $\{X, ZZ, Z, XX, YY\}$, \textbf{o}. $\{X, ZZ, Y, XX, YY\}$, and \textbf{p}. $\{X, Y, Z, XX, YY, ZZ\}$
		}
		\label{supp_fig_2}
	\end{figure}
	
	\section{Quantum circuit ansatzes}
	
	We depict quantum circuit ansatzes used in all our experiments and simulations. We used two different kind of ansatzes: the Alternating Layered Ansatz (ALA) and the YY ansatz.
	
\begin{figure}[H]
		\centering
		\includegraphics[width = 0.7\textwidth]{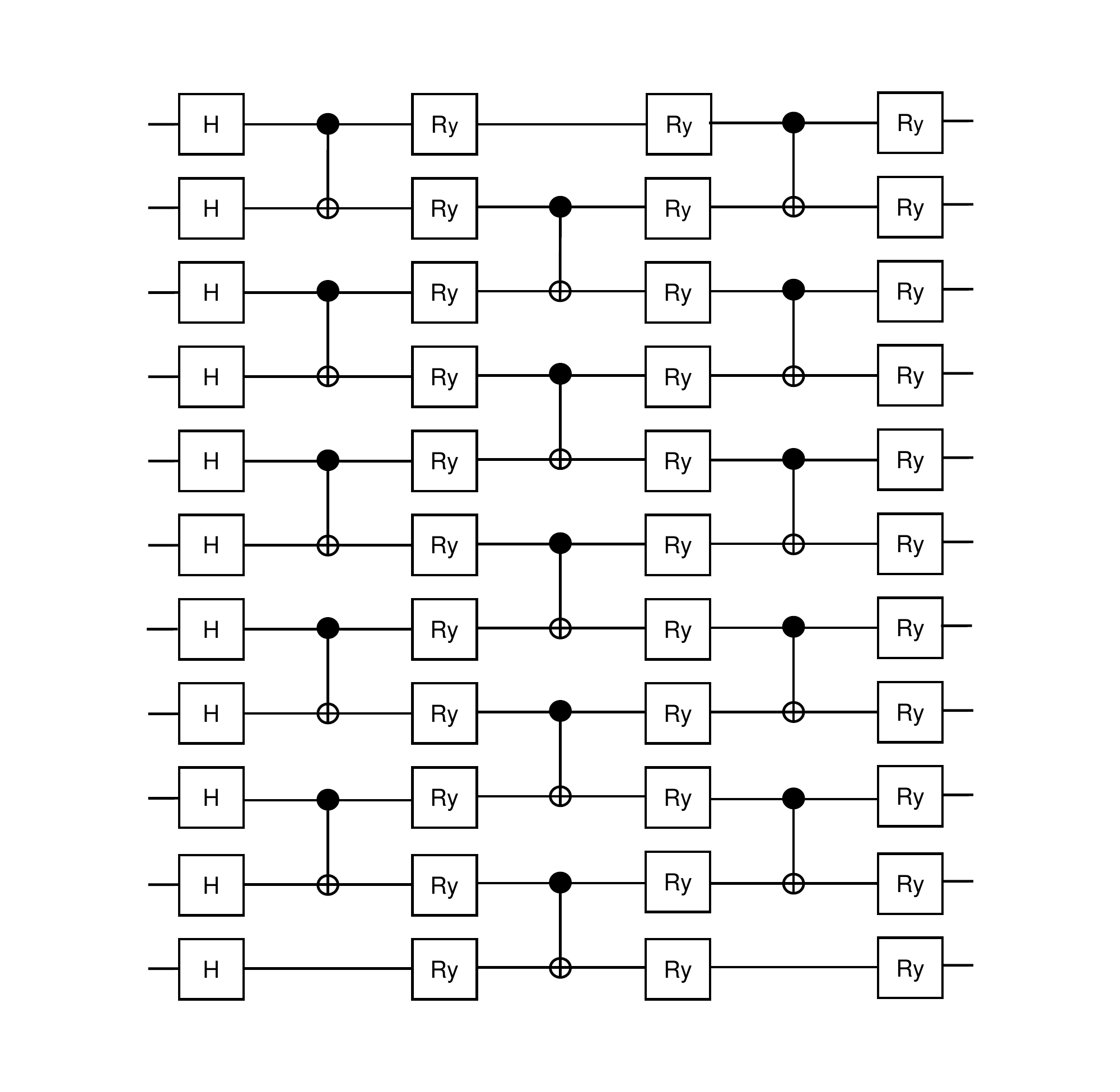}
		\vspace{-5mm}
		\caption{
			11-qubit 3 layer ALA ansatz
		}
		\label{supp_fig_3}
	\end{figure}

\begin{figure}[H]
		\centering
		\includegraphics[width = 0.7\textwidth]{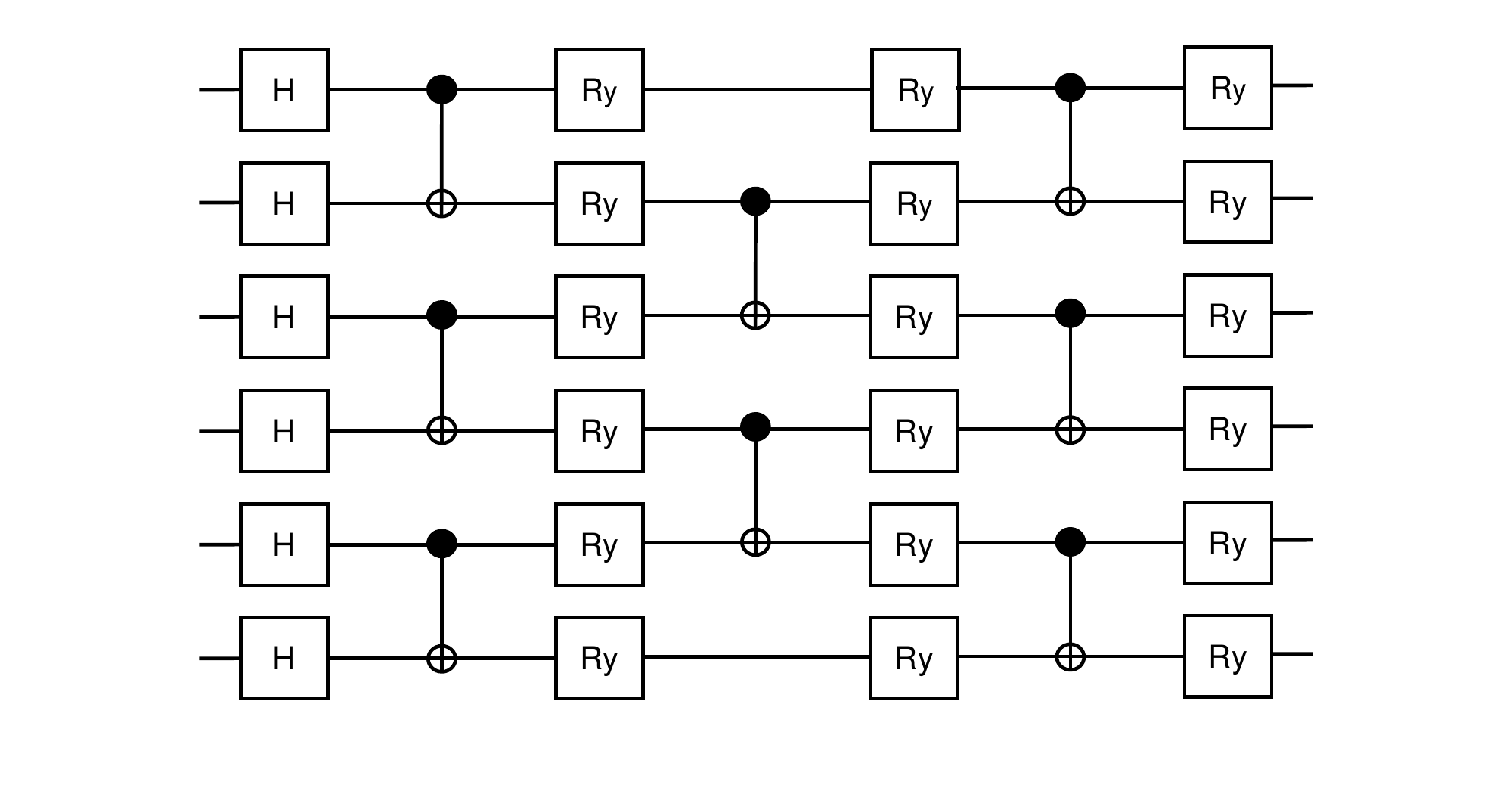}
		\vspace{-5mm}
		\caption{
			6-qubit 3 layers ALA ansatz
		}
		\label{supp_fig_4}
	\end{figure}

	ALA was first introduced in reference \cite{cerezo2021cost} and was shown to be expressive \cite{nakaji2021expressibility}. However, we altered the ALA slightly for our purposes. We empirically observed that replacing the first layer of Ry gates with a layer of Hadamard gates improved optimization speed and the chance of converging to global minimums. We thus replaced the first layer of Ry gates with Hadamard gates. Appenidx Figure \ref{supp_fig_3} and \ref{supp_fig_4} are 11-qubit 3 layer ALA and 6-qubit 3 layer ALA used to run experiments and simulations shown in main text Figure 2a and 2b respectively.
	
	11-qubit 3 layer ALA ansatz was used for experiments and simulations that generated main text Figure 2a and 2b, Appendix Figure \ref{supp_fig_6}, Appendix Figure \ref{supp_fig_10}, and Appendix Figure \ref{supp_fig_12}a.
	8-qubit 3 layer ALA ansatz was used for simulations that generated Appendix Figure \ref{supp_fig_11}.
	6-qubit 3 layer ALA ansatz was used for experiments and simulations that generated main text Figure 2c and 2d, Appendix Figure \ref{supp_fig_7}, Appendix Figure \ref{supp_fig_10}, and Appendix Figure \ref{supp_fig_12}b.
	
	YY ansatz is hand-designed by our group and has shown to be effective in expressing the ground state of the 1D-TFIM Hamiltonian. 11-qubit 1 layer YY ansatz is depicted in Appendix Figure \ref{supp_fig_5}b. To add another layer to the YY ansatz, we simply repeat all the CNOT and the $R_{y}$ gates in Appendix Figure \ref{supp_fig_5}. We used 11-qubit 2 layer YY ansatz in simulations shown in main text Figure 3a, 3b, and 3c.
	
	\begin{figure}[H]
		\centering
		\includegraphics[width = \textwidth]{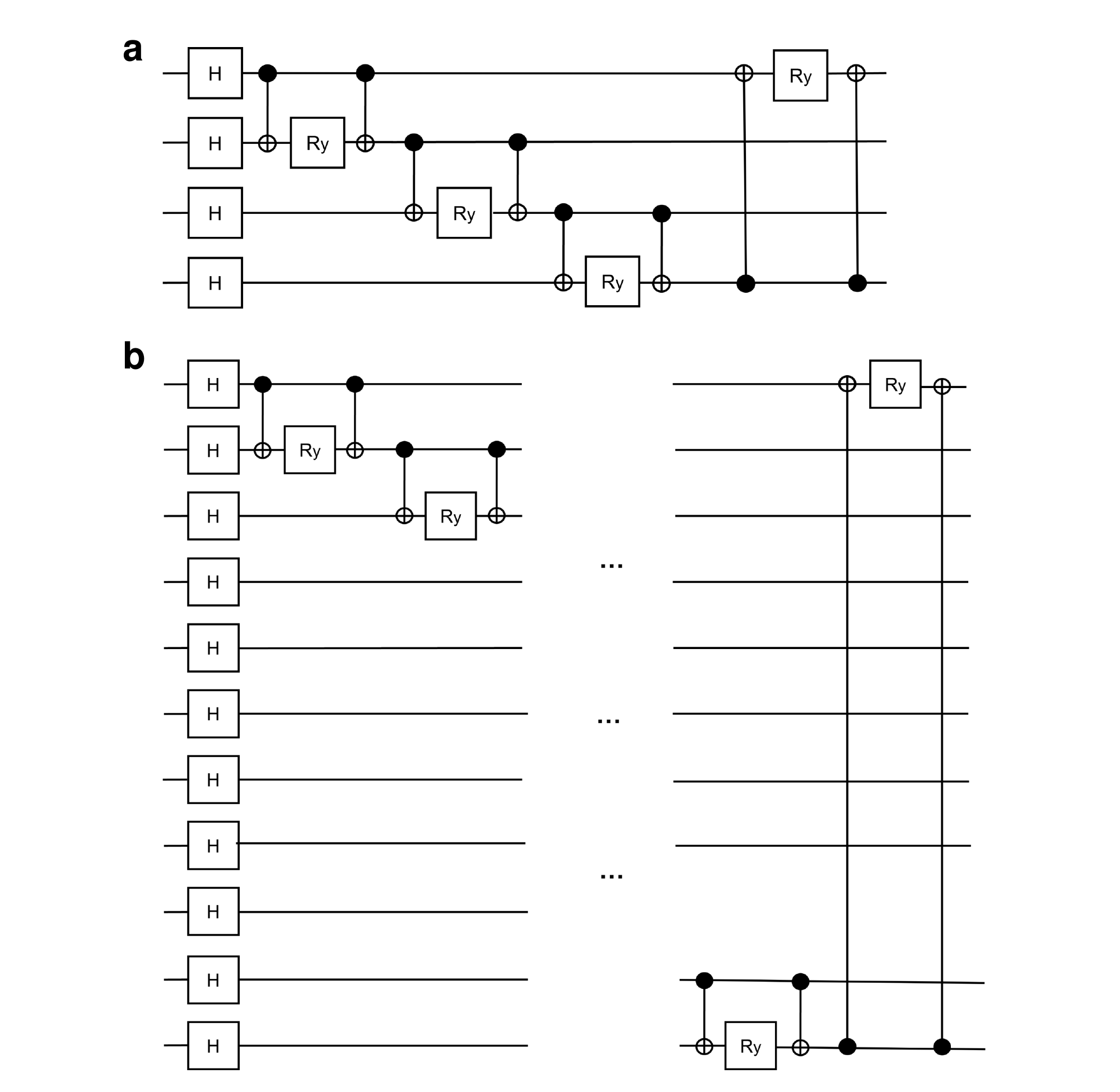}
		\caption{ Depiction of YY ansatz. \textbf{a}. 4-qubit 1 layer YY ansatz. \textbf{b}. 11-qubit 1 layer YY ansatz, generalized from the 4-qubit 1 layer YY ansatz.
		}
		\label{supp_fig_5}
	\end{figure}
	
	\section{Energy, HR distance, and variance: raw data from IonQ device}
	\label{sec:section6}
	
	We show HR distances and Hamiltonian variances measured using lists of parameters from simulated VQE, without any moving averages. We further show how the energy measured with IonQ device varies with respect to VQE iterations. To clarify, the energy data shown in Appendix Figure \ref{supp_fig_6} and \ref{supp_fig_7} are not the simulated energy that we used to optimize our simulated VQE, but energy measured with IonQ device only using the parameters from the simulated VQE.
	
	\begin{figure}[H]
		\includegraphics[width = \textwidth]{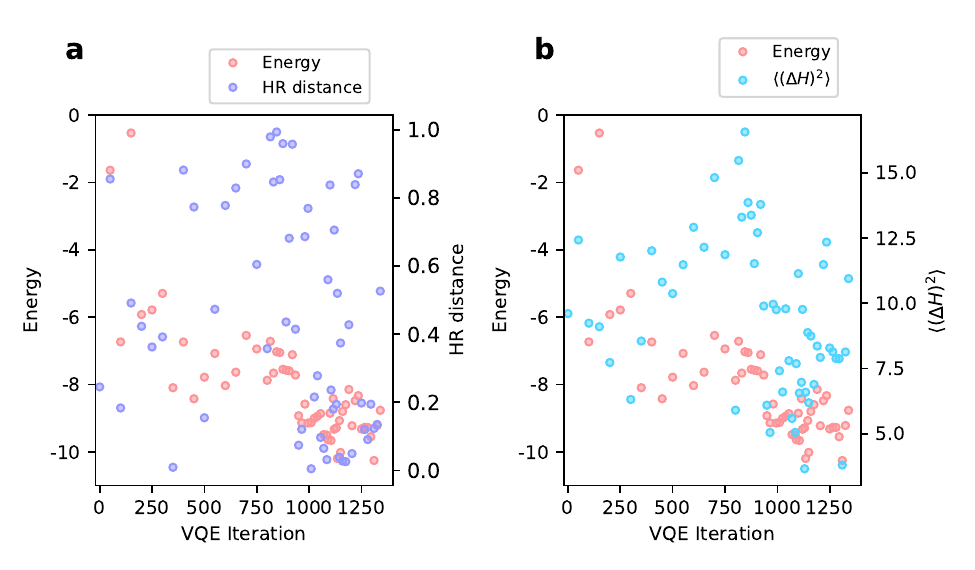}
		\caption{ HR distance and Hamiltonian variance measured with IonQ device using lists of parameters sampled from the simulated VQE for the 11-qubit 1D-TFIM Hamiltonian. Both subplots further show how energy measured with IonQ device changes with simulated VQE iterations. We sampled the lists of parameters from the simulated VQE and measured the HR distances (Hamiltonian variances) and energies with an IonQ's device for \textbf{a} (\textbf{b}). Light red points are energies for both \textbf{a} and \textbf{b}. Blue points in \textbf{a} are HR distances, and the light blue points in \textbf{b} are variances of the Hamiltonian.
		}
		\label{supp_fig_6}
	\end{figure}
	
	\begin{figure}[H]
		\includegraphics[width = \textwidth]{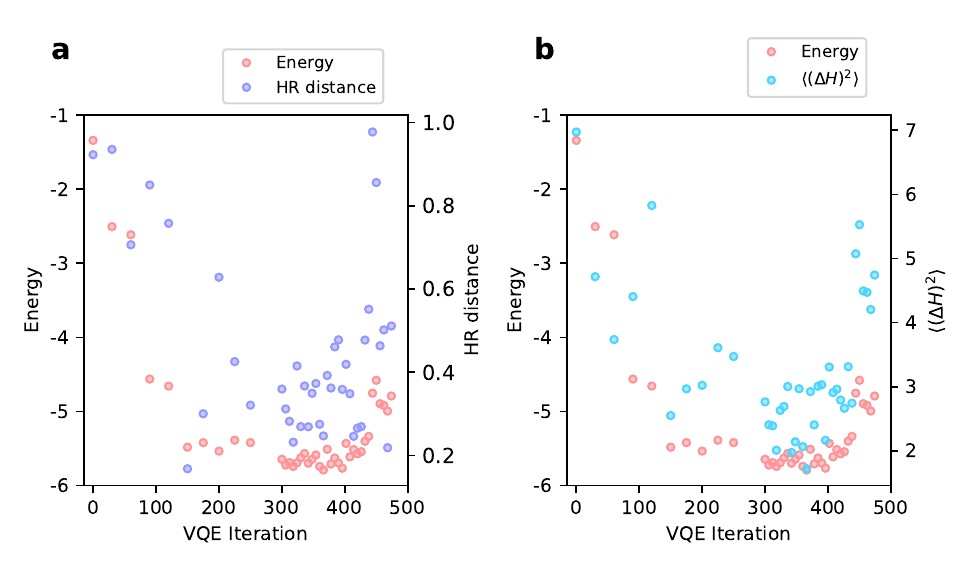}
		\caption{HR distance, Hamiltonian variance, and energy measured with IonQ device using the list of parameters sampled from the simulated VQE for the 6-qubit J1-J2 TFIM Hamiltonian. Color conventions are same as the Appendix Figure \ref{supp_fig_6}, except that all data points are for the 6-qubit $J_{1}$-$J_{2}$ TFIM Hamiltonian.
		}
		\label{supp_fig_7}
	\end{figure}
	
	To obtain the lists of parameters to use when measuring HR distances and variances of the Hamiltonian, we simulate VQE by first randomly initializing all the parameters and then updating parameters with the IMFIL optimizer (discussed in \ref{sec:section7}). We used 4,000 and 10,000 shots to measure the energy from simulated VQE for the 11-qubit 1D-TFIM Hamiltonian and the 6-qubit $J_{1}$-$J_{2}$ TFIM Hamiltonian respectively. The number of shots was chosen by simulating the VQE with varying number of shots and by observing that too few in shots caused VQEs to converge to non-ideal local minimums. We further did not want to simulate with an extremely high number of shots so that we can reasonably run the simulated VQE in an actual quantum hardware if we were to. We did not add any depolarization noise during our VQE simulations because our goal was simply to obtain lists of parameters that approximately got the ansatz close enough to the ground state. We then sampled indices from VQE iterations, with more points sampled near the end of iterations to show how the HR distance changed near VQE convergence.

	Appendix Figure \ref{supp_fig_7}a and \ref{supp_fig_7}b show how the energy measured with IonQ device increases near the end of the VQE iteration (after around $430^{th}$ iterations), even though the VQE progresses for the 6-qubit $J_{1}$-$J_{2}$ TFIM Hamiltonian. This discrepancy exists because we used the simulated VQE instead of running the VQE on the IonQ's device and thus the simulated VQE ended up finding states with non-ideal local minimums. We further see how the HR distance is indicative of VQE finding states in non-ideal local minimum; in Appendix Figure $\ref{supp_fig_7}$a, the HR distance increases near the end of the VQE optimization, as the simulated VQE found states with energies that are not fully minimized.
	
	\section{Variational Quantum Eigensolver Optimization}
	\label{sec:section7}

	VQE is a hybrid quantum computing algorithm that can be used to solve for the ground state energy of quantum systems. VQE relies on the Rayleigh-Ritz principle shown in equation \ref{eq:5} \cite{Tilly_2022}. This states that the ground state energy will always be less than the expectation value of the Hamiltonian, $\hat{H}$ over the probability density of the parameterized wavefunction, $|\psi(\theta)\rangle$.
	\begin{equation}
		\begin{aligned}
			E_{0} \leq \frac{\langle\psi(\theta)|\hat{H}| \psi(\theta)\rangle}{\langle\psi(\theta)|\psi(\theta)\rangle}
		\end{aligned}
		\label{eq:5}
	\end{equation}
	
	So by optimizing the parameters of the wave function, the ground state energy and corresponding ground state wavefunction can be found. With a prepared state, $|\mathbf{0}\rangle$, and an ansatz, $U^{\dagger}(\boldsymbol{\theta})$, we can rewrite the parameterized state to get equation \ref{eq:6}.

	
	\begin{equation}
		\begin{aligned}
			E_{s\mathrm{VQE}}=\min _{\boldsymbol{\theta}}\langle\mathbf{0}|U^{\dagger}(\boldsymbol{\theta}) \hat{H} U(\boldsymbol{\theta})| \mathbf{0}\rangle
		\end{aligned}
		\label{eq:6}
	\end{equation}
	
	Equation \ref{eq:6} shows as we optimize our ansatz, we can minimize the expected value of our given Hamiltonian to find our ground state energy and the parameters with which we can prepare our ground state wavefunction $|\psi_{gs}(\theta)\rangle = U(\boldsymbol{\theta})| \mathbf{0}\rangle$. For our simulations, to calculate the energy of a given ansatz we can simply take the matrix multiplication of our Hamiltonian with our parameterized state, and then multiply the conjugate of the parameterized state times the result. 
	
	To optimize our VQE simulations, we used ImFil, Implicit Filtering. Implicit Filtering is an algorithm used to solve bound-constrained optimization problems \cite{kelley2011implicit}. ImFil uses a stencil, or coordinate search, followed by interpolation to approximate the gradient. This algorithm is built for problems with local minima caused by high-frequency, low-amplitude noise and with an underlying large-scale structure that is easily optimized. The initial clusters are controlled by the boundaries, meaning ImFil is relatively insensitive to the initial point and allows ImFil to escape from local minima \cite{optimizer_NISQ}. The step sizes ImFil uses also decrease after each iteration allowing for more accurate convergence to the lowest value. These factors make ImFil a good choice of optimizer for our purposes as we do not know a good initial point and need close convergence to the ground state energy and wavefunction. 
	
	\section{Effect of depolarization and shot noise on the HR distance}
	
	We analyze how depolarization noise affects the value of the HR distance and further investigate how the correlation between the HR distance and the fidelity depends on the depolarization noise. Definition of the depolarization noise is written in equation (\ref{eq:7s}) \cite{nielsen2001quantum}. 
	
	\begin{equation}
		\begin{aligned}
			\mathcal{E}(\rho) = \frac{pI}{d} + (1-p)\rho
		\end{aligned}
		\label{eq:7s}
	\end{equation}
	
	$d$ is the dimension of the system that scales as $2^{n}$, where $n$ is the number of qubits. $\rho$ is the initial state before applying the depolarization noise. $p$ is the value of the depolarization noise. $I$ is the identity operator for a $d$ dimensional system, and lastly $\mathcal{E}(\rho)$ is the resulting state after the depolarization noise is applied.
	
	To analyze how depolarization noise affects the ground state of the 11-qubit 1D-TFIM and the 6-qubit $J_{1}$-$J_{2}$ TFIM, we incrementally added depolarization noise to ground states of both systems and measured corresponding HR distances and fidelities. Results are shown in Appendix Figure \ref{supp_fig_8}a and \ref{supp_fig_8}b for 11-qubit 1D-TFIM and 6-qubit $J_{1}$-$J_{2}$ TFIM respectively.

	\begin{figure}[H]
		\includegraphics[width = \textwidth]{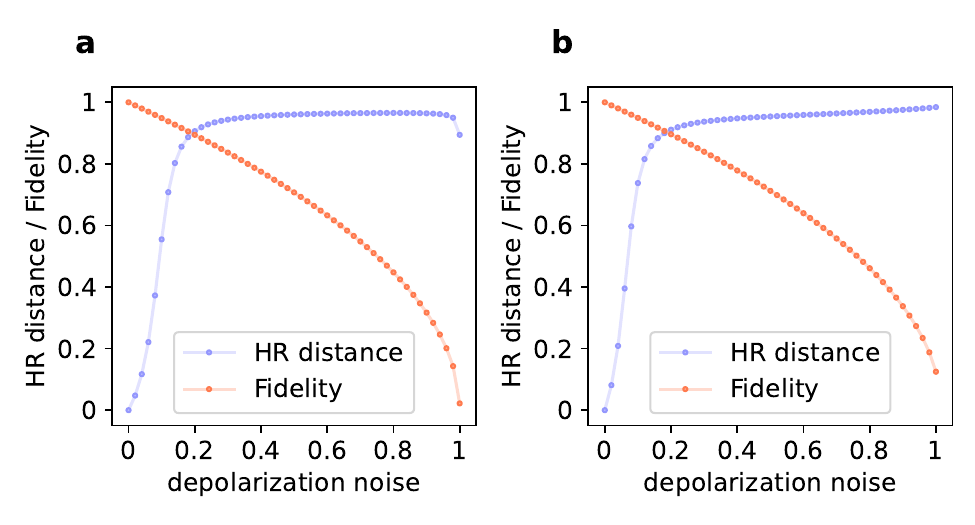}
		\caption{
			HR distance and fidelity measured for the 11-qubit 1D-TFIM (\textbf{a}) and for the 6-qubit $J_{1}$-$J_{2}$ TFIM (\textbf{b}), while incrementally increasing the depolarization noise.
		}
		\label{supp_fig_8}
	\end{figure}

	For both models, we see negative correlations between HR distances and fidelities as we vary depolarization noise. However, we further see how a certain value of fidelity does not necessarily correspond to a certain value of HR distance across different models. For example, in Appendix Figure \ref{supp_fig_2}a, fidelity value of $0.9$ corresponds to HR distance value of approximately $0.75$, but for Appendix Figure \ref{supp_fig_8}a, fidelity value of $0.9$ corresponds to HR distance value of approximately $0.9$. Thus, the HR distance metric should not be used as an equivalent substitute of the fidelity but be used as an another method to diagnose the VQE solution. 
	
\begin{figure}[H]
		\includegraphics[width = \textwidth]{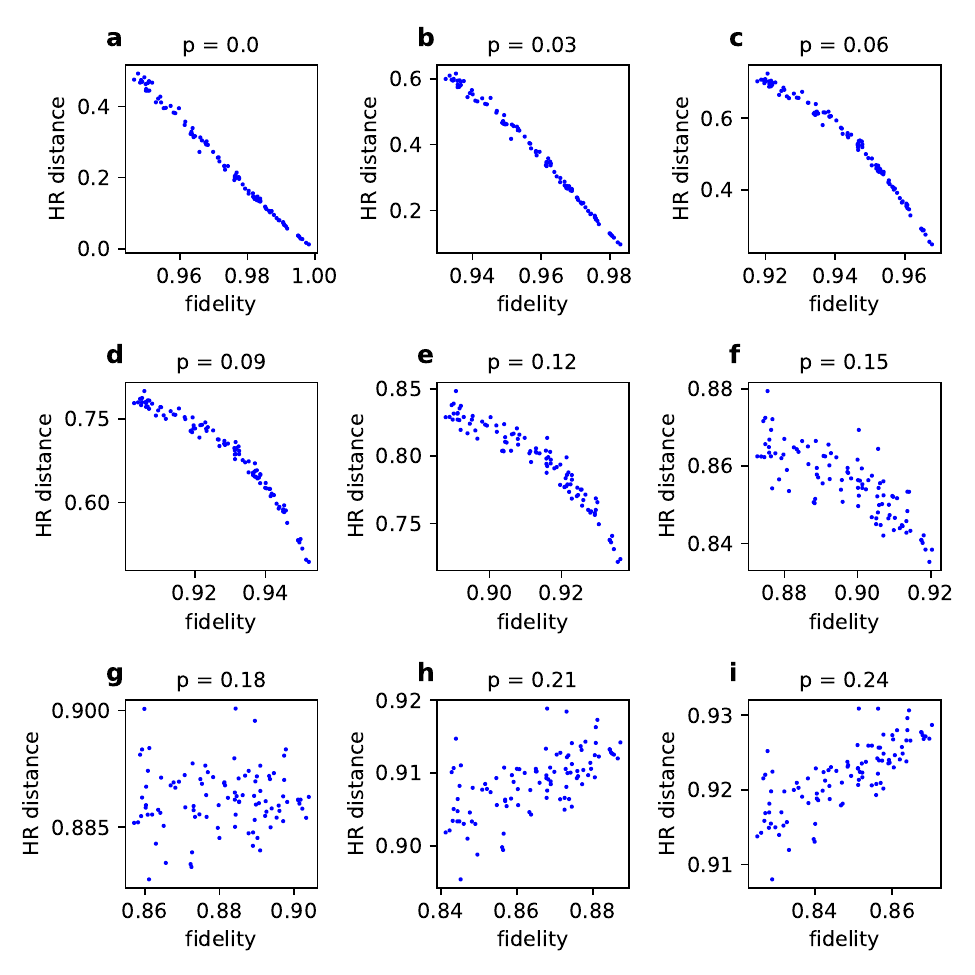}
		\caption{
			HR distance and Fidelity measured for randomly perturbed wavefunctions described in \ref{section:B} using different values of depolarization noise for the 11 qubit 1D-TFIM. The depolarization noise value for each of the subplot is listed as p. 
		}
		\label{supp_fig_9}
	\end{figure}

	Appendix Figure \ref{supp_fig_9} shows the effect of depolarization noise on the negative correlation between the HR distance and the fidelity for the 11 qubit 1D-TFIM. We used randomly perturbed wavefunctions described in \ref{section:B} for the 11 qubit 1D-TFIM and incrementally added depolarization noise following equation (\ref{eq:7s}) and plotted the fidelity and the HR distance. We used operators only in the Hamiltonian: $\{\sum_{i=1}^{N} \sigma_i^{x}, \sum_{i=1}^{N-1} \sigma_i^{z}\sigma_{i+1}^{z}\}$ for Hamiltonian Reconstruction.
	
	When the probability of depolarization noise is more than 0.1 ($p > 0.1$), we start to lose the negative correlation between the HR distance and the fidelity drastically. When $p \geq 0.18$. we do not see any correlation. We further see how the range of HR distance values changes as we add more depolarization noise.
	
	Observing how a certain level of depolarization noise causes to lose the correlation, we can infer that depolarization noise from 1-qubit and 2-qubit gates can introduce enough depolarization noise to lose the correlation between the HR distance and the fidelity. This shortcoming indicates how, for a certain quantum hardware with fixed 1-qubit and 2-qubit gate fidelities, there is a limitation on how many gates that we can apply before the HR distance loses its negative correlation with the fidelity.
	
	Also, note how the HR distance values change as we inject more depolarization noise. From this, we can further infer that depolarization noise from 1-qubit and 2-qubit gates should also give different values of HR distances at VQE convergence, which in turn indicates that the HR distance at the convergence is hardware dependent. Thus, it is necessary to know the the 1-qubit gate and 2-qubit gate fidelities in order to do an analysis on what value of HR distance indicates an optimal VQE solution or not for any particular quantum hardware.
	
	\begin{figure}[H]
		\centering
		\includegraphics[width = \textwidth]{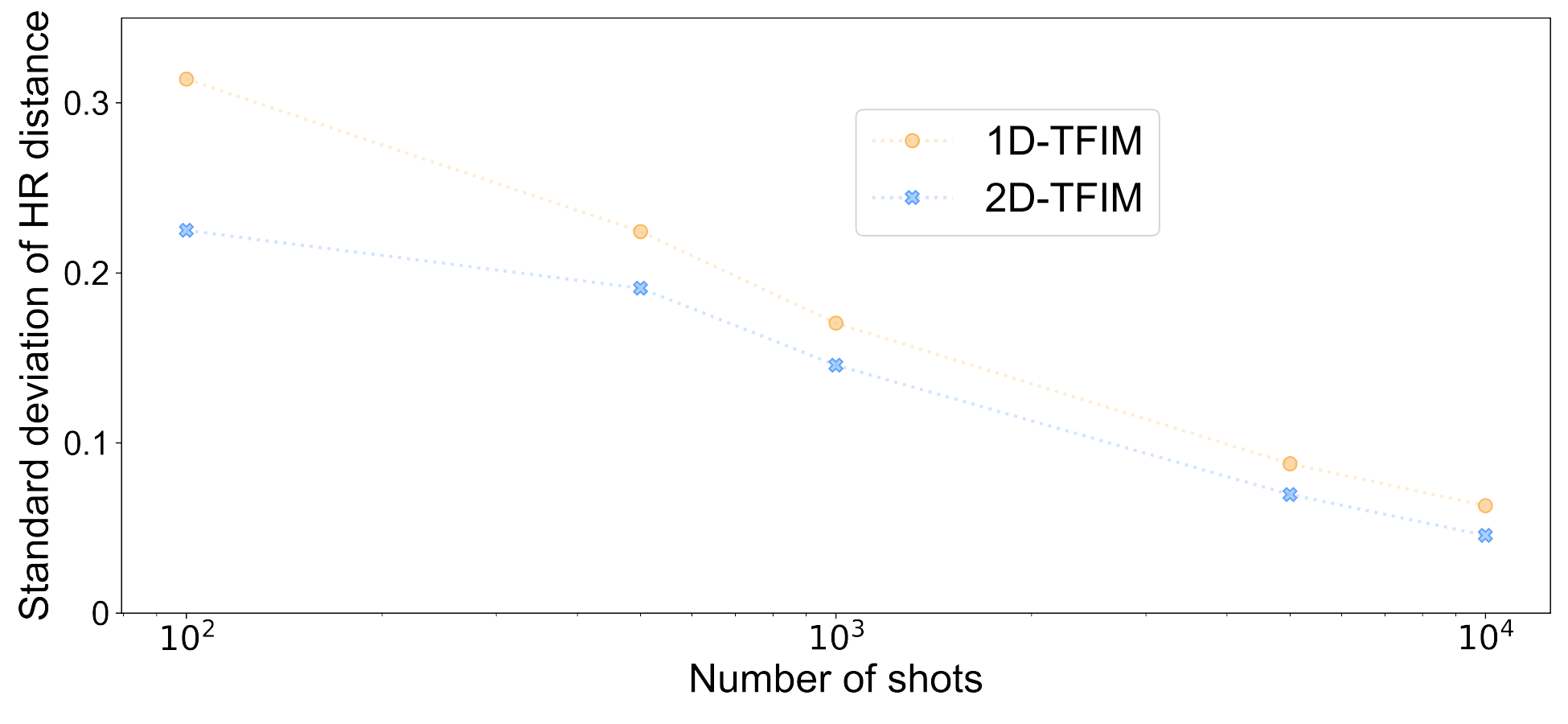}
		\caption{Dependence of the HR distance on the number of shots used in measurements for the 11-qubit 1D-TFIM and the 6-qubit $J_{1}$-$J_{2}$ TFIM. All simulations were run using the ALA ansatzes in Appendix E with depolarization noise values equal to the reported average IONQ harmony gate infidelities.}
		\label{supp_fig_10}
	\end{figure}
	
	Appendix Figure \ref{supp_fig_10} shows, in simulation, the standard deviation of the HR distance as a function of the number of shots used for measurement. The ansatz used for this simulation was the ALA (the same as the one used for the experiments in main text figures 2b and 2d). One can see that a large number of shots ($\sim 10^{4}$) is required to have a reasonable signal-to-noise ratio (SNR) for the HR distance (the mean value of the HR distance in Figure 3e was $\sim$ 0.2 for both models).
	
	\section{Robustness of the HR distance to perturbations}
	In this section, we show how HR distance can be noise-robust with small perturbations in measuring the covariance matrix. Proof in this section is derived from \cite{qi2019determining}.
	
	Consider the state $\ket{\Psi}$, one of the output states of VQE. From the output state $\ket{\Psi}$, we can measure the covariance matrix $Q[\Psi, H]$, where $Q[\Psi, H]_{i,j} = \frac{1}{2}(\langle H_{i}H_{j} \rangle + \langle H_{j}H_{i} \rangle) - \langle H_{j}\rangle \langle H_{j}\rangle$
	
	Let us say that the covariance matrix is perturbed, due to the noise in the quantum ansatz used in VQE (noise is incorporated in $\ket{\Psi}$).
	
	Let the perturbed covariance matrix as $Q'$ where $Q' = Q + \epsilon \Delta Q$
	
	Let the eigenvalues of the unperturbed covariance matrix be $\sigma_{1} \leq \sigma_{2} ... \leq \sigma_{N}$ with their corresponding eigenvectors as $v_{1}, v_{2}, ... , v_{N}$. Let $\sigma_{1}' \leq \sigma_{2}' ... \leq \sigma_{N}'$ be the eigenvalues of the perturbed covariance matrix and their corresponding eigenvectors as $v_{1}', v_{2}', ... , v_{N}'$
	
	Because the eigenvector corresponding to the lowest eigenvalue of the unperturbed covariance matrix is the reconstructed Hamiltonian from it, let us denote it as $H_{R} (= v_{1})$. Similarly, let us denote that of $Q'$ as $H_{R}' (= v_{1}')$.
	
	Our goal is to bound $\abs{\norm{H_{orig} - H_{R}'}_{2} - \norm{H_{orig} - H_{R}}_{2}}$, which is the difference between the HR distance from the perturbed covariance matrix and that from the unperturbed covariance matrix.
	
	Consider the triangle formed by $H_{orig}, H_{R}, \text{ and } H_{R}'$, where $H_{orig}$ is the original Hamiltonian. We use the triangle inequality written below:
	\begin{align*}
		\norm{H_{orig} - H_{R}}_{2} + \norm{H_{R} - H_{R}'}_{2} \geq \norm{H_{orig} - H_{R}'}_{2} \\
		\norm{H_{R} - H_{R}'}_{2} \geq \norm{H_{orig} - H_{R}'}_{2} - \norm{H_{orig} - H_{R}}_{2} \\
		\therefore \abs{\norm{H_{orig} - H_{R}'}_{2} - \norm{H_{orig} - H_{R}}_{2}} \leq \norm{H_{R} - H_{R}'}_{2} \\
	\end{align*}
	
	Further, using cosine equality of the triangle:
	
	\begin{align*}
		\norm{H_{R} - H_{R}'}_{2}^{2} &= \norm{H_{R}}_{2}^{2} + \norm{H_{R}'}_{2}^{2} - 2\norm{H_{R}}_{2}\norm{H_{R}'}_{2}cos(\theta_{(H_{R}, H_{R}')}) \\
		&= 2 - 2cos(\theta_{(H_{R}, H_{R}')}) (\because H_{R} \text{ and } H_{R}' \text{ both normalized to 1}) \\
		&= 2 - 2H_{R} \cdot H_{R}' (\because cos(\theta_{(H_{R}, H_{R}')}) = \frac{H_{R} \cdot H_{R}'}{\norm{H_{R}}_{2}\norm{H_{R}'}_{2}} = H_{R} \cdot H_{R}') \tag{A}\label{eq:A}
	\end{align*}

	From perturbation theory, we can further relate $H_{R}$, $H_{R}'$ by
	
	\begin{align*}
		c H_{R}' &= H_{R} + \epsilon H^{a} + \epsilon^{2}H^{b} + ... (\text{c is normalization constant})\\
		\sigma_{1}' &= \sigma_{1} + \epsilon \sigma_{1}^{a} + \epsilon^{2} \sigma_{1}^{b} + ... \\
		Q'H_{R}' &= \sigma_{1}'H_{R}' = (Q + \epsilon\Delta Q)(H_{R} + \epsilon H^{a} + \epsilon^{2} H^{b} + ... )\\
		&= (\sigma_{1} + \epsilon \sigma_{1}^{a} + \epsilon^{2} \sigma_{1}^{b} + ... )( H_{R} + \epsilon H^{a} + \epsilon^{2}H^{b} + ... ) \\
		\text{With respect to }& \epsilon,\\
		\text{ zeroth-order term: } & QH_{R} = \sigma_{1}H_{R} \\
		\text{ first-order term: } & QH^{a} + \Delta Q H_{R} = \sigma_{1}^{a}H_{R} + \sigma_{1}H^{a} \\
		& (Q - \sigma_{1} I) H^{a} = (\sigma_{1}^{a}I - \Delta Q)H_{R}
	\end{align*}
	
	Let $H^{a} = \sum_{i=1}^{N} c_{i}v_{i}$, then from the first-order term, we can conclude that:
	
	\begin{align*}
		& cH_{R}' = H_{R} + \epsilon \sum_{i=2}^{N}c_{i}v_{i}, \text{ where } c_{i} = -\frac{v_i^{T}\Delta Q H_{R}}{\sigma_{1} - \sigma_{i}} \text{ (to first order approximation)}\\
		& H_{R}' = \frac{H_{R} + \epsilon \sum_{i=2}^{N}c_{i}v_{i}}{\sqrt{1 + \sum_{i=2}^{N}\epsilon^{2}c_{i}^{2}}}, (\text{normalizing } H_{R}')\\
		& \frac{1}{\sqrt{1 + \sum_{i=2}^{N}\epsilon^{2}c_{i}^{2}}} \geq 1 - \sum_{i=2}^{N}\epsilon^{2}c_{i}^{2}/2, ( \epsilon \text{ is small}) \\
		& H_{R} \cdot H_{R'} = \frac{1}{\sqrt{1 + \sum_{i=2}^{N}\epsilon^{2}c_{i}^{2}}} \geq 1 - \sum_{i=2}^{N}\epsilon^{2}c_{i}^{2}/2 \tag{B}\label{eq:B}
	\end{align*}
	
	Note that $\epsilon^{2}\sum_{i=2}^{N}c_{i}^{2} \leq \frac{\epsilon^{2}}{\sigma_{2}^{2}}\norm{\Delta Q}^{2}$...  (C), where the $\sigma_{2}$ is the second smallest eigenvalue and norm applied to $\Delta Q$ is the 2-norm for matrix. Here, we leave out the higher order $\epsilon$ terms.
	
	Therefore, from $(A), (B), \text{ and } (C)$, the difference between the HR distance from the perturbed covariance matrix and that from the unperturbed covariance matrix is   $\abs{\norm{H_{orig} - H_{R}'}_{2} - \norm{H_{orig} - H_{R}}_{2}} \leq \norm{H_{R} - H_{R}'}_{2} \leq  \sqrt{\sum_{i=2}^{N}\epsilon^{2}c_{i}^{2}} \leq  \frac{\epsilon}{\sigma_{2}}\norm{\Delta Q}$.
	
	By bounding the HR distance in a small perturbation limit, we show how HR distance can potentially be noise-robust, especially when the second largest eigenvalue of the covariance matrix is large enough.
	
	\section{How the gap between the ground state and the excited state affects the HR distance}
	
	\begin{figure}
		\includegraphics[width = 1.0\textwidth]{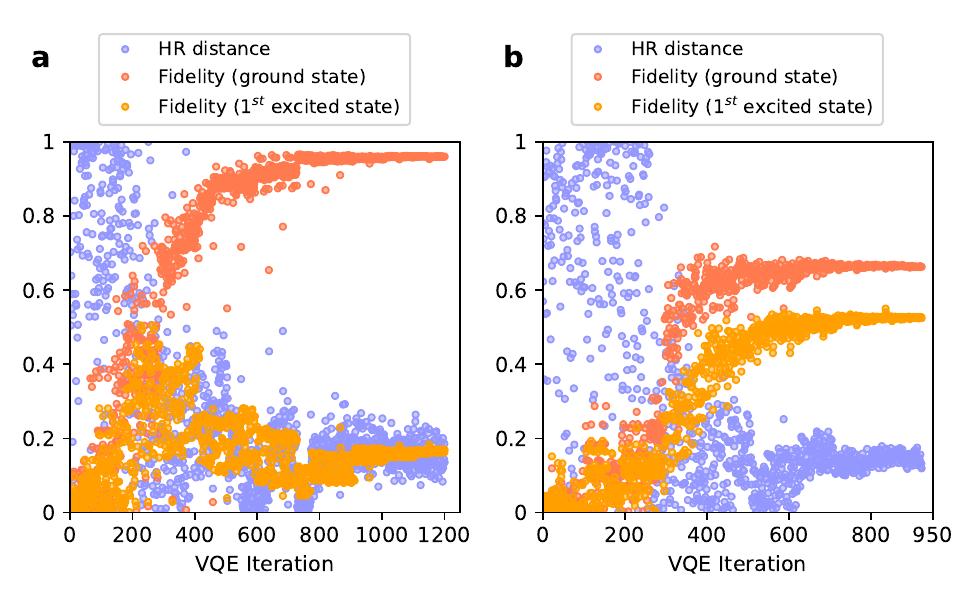}
		\caption{ VQE ran with \textbf{a)} $J = 0.5$ and \textbf{b)} $J = 1$ for 8 qubit 1-D Transverse Field Ising model. Blue points are HR distances measured for each iteration of VQE optimization. Red points are fidelities between the output state from each iteration of VQE optimization and the ground state. Lastly, orange points are fidelities between the output state from each iteration of VQE optimization and the first excited state.
		}
		\label{supp_fig_11}
	\end{figure}
	
	We discuss how the gap between the energy of the ground state ($E_{gst}$) and the energy of the first excited state ($E_{fst}$) affects the HR distance. We quantify the magnitude of the gap as the difference between the energy of the ground state and the energy of the first excited state over the energy scale of the system $J$ ($\text{gap} = |E_{gst} - E_{fst}|/|J|$). Here, we consider the 8-qubit 1D-TFIM Hamiltonian but with periodic coupling described in (\ref{eq:4}).

	\begin{equation}
		\begin{aligned}
			H_{tfim} = \sum_{i=1}^{N} \sigma_i^{x} + J \cdot \sum_{i=1}^{N} \sigma_{i}^{z}\sigma_{(i+1)\%N}^{z} 
			\label{eq:4}
		\end{aligned}
	\end{equation}

	$\sigma_{i}^{x}$ and $\sigma_{i}^{z}$ correspond to the pauli-x and pauli-z operators on the $i^{th}$ spin. $J$ is the coupling strength of the nearest-neighbor spin interaction for the 1D-TFIM Hamiltonian. $N$ is the number of qubits in the system. We added the periodic coupling because when $J = 1$, the gap between $E_{gst}$ and $E_{fst}$ is significantly smaller than that of the 1D-TFIM Hamiltonian without the periodic coupling. Appendix Figure \ref{supp_fig_11} shows how the HR distance, the fidelity with the ground state, and the fidelity with the first excited state change during the simulated VQE optimization for the 1D-TFIM Hamiltonian with $J = 0.5$ (Appendix Figure \ref{supp_fig_11}a) and with $J = 1$ (Appendix Figure \ref{supp_fig_11}b). During simulations, we did not add any depolarization noise, but for each iteration of the VQE optimization, we used 10,000 shots to measure the energy and 4,000 shots to measure the HR distance. Furthermore, we used the operators only in the Hamiltonian, $\sum_{i=1}^{N} \sigma_i^{x} \text{ and }  \sum_{i=1}^{N} \sigma_{i}^{z}\sigma_{(i+1)\%N}^{z}$, to construct the set of operators for the Hamiltonian reconstruction. All fidelity measurements were done without shot noise and depolarization noise.
	
	When $J = 0.5$, the gap is equal to $|E_{gst} - E_{fst}|/|J| = |-8.509 - (-7.508)|/|0.5| \approx 2.00$ and when $J = 1$, the gap is equal to $|E_{gst} - E_{fst}|/|J| = |-10.25 - (-10.05)|/|1| \approx 0.20$. When the gap is relatively small ($\approx 0.20$) HR distance also has 
	a significant negative correlation with the fidelity with the first excited state in contrast with when there is a large gap ($\approx 2.00$). We thus concluded that we can not substitute the fidelity with the HR distance especially when the gap between the energy of the ground state and the energy of the first excited state
	is small, as the HR distance also becomes negatively correlated with the fidelity with the first excited state.
	
	\section{Hamiltonian Variance measurement}
	\label{sec:section5}
	
	Measuring the variance of the Hamiltonian ($\langle (\Delta H)^{2} \rangle$) \cite{zhang2022variational} has been suggested as an alternative cost function to use for VQE optimization. To compare it to the HR distance, we perform the following analysis. Using the same simulated VQE used in Figure 2 of the main text, we obtain the variances of the Hamiltonian, instead of the HR distances, using the same measurements from the IonQ's device. We further simulate the variance of the Hamiltonian with shot noise and depolarization noise, using 4,000 shots (10,000 shots) to measure each observable for the 11-qubit 1D-TFIM Hamiltonian (6-qubit $J_{1}$-$J_{2}$ TFIM Hamiltonian) with 1-qubit and 2-qubit depolarization noise values set as IonQ device's 1-qubit and 2-qubit average gate fidelities (0.0065 and 0.0398 respectively).
	
\begin{figure}[H]
		\includegraphics[width = \textwidth]{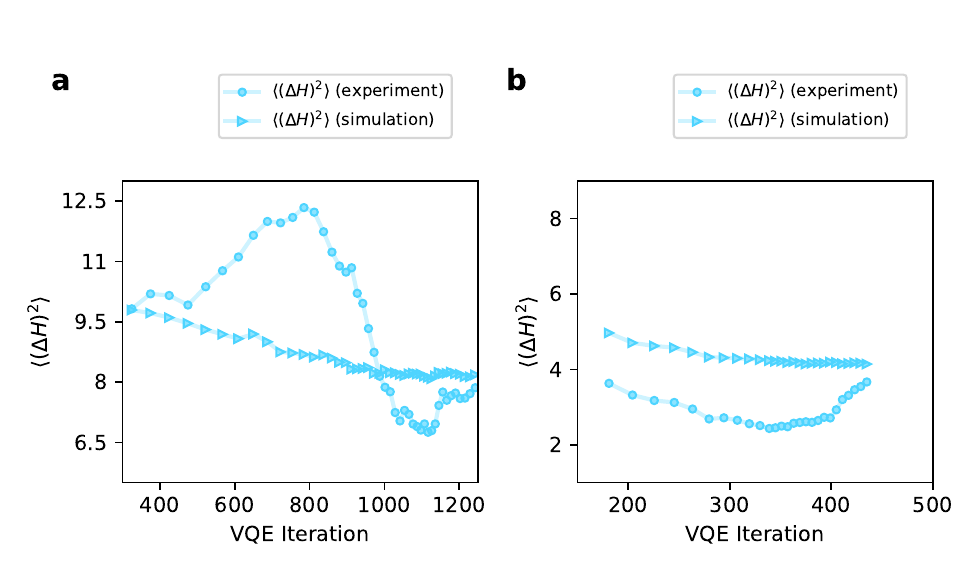}
		\caption{Hamiltonian variance ($ \langle (\Delta H)^{2} \rangle$) from both experiment (with IonQ device) and simulation, using the list of parameters from the simulated VQE. A 14-point moving average window was used to better show the correlation between the Hamiltonian variance and the fidelity. \textbf{a)} simulated Hamiltonian (light-blue circle) variance and Hamiltonian variance measured with IonQ device (light-blue triangle) for the 11-qubit 1D-TFIM Hamiltonian. \textbf{b)} simulated Hamiltonian variance (light-blue circle) and Hamiltonian variance measured with IonQ device (light-blue triangle) for the 6-qubit $J_{1}$-$J_{2}$ TFIM Hamiltonian.
		}
		\label{supp_fig_12}
	\end{figure}

	Since we only used 4,000 shots (10,000 shots for the 6-qubit $J_{1}$-$J_{2}$ TFIM Hamiltonian) during our experiment, we took a 14-point moving average of variances to better show the trajectory of the data. Raw data is shown in Appendix Figure \ref{supp_fig_6} and \ref{supp_fig_7}.
	Appendix Figure \ref{supp_fig_12}a and \ref{supp_fig_12}b show how the simulated variance of the Hamiltonian and the variance of the Hamiltonian measured with IonQ device vary as VQE progresses for the 11-qubit 1D-TFIM Hamiltonian and the 6-qubit $J_{1}$-$J_{2}$ TFIM Hamiltonian respectively. 
	
	For both Appendix Figure \ref{supp_fig_12}a and \ref{supp_fig_12}b, we see that the variance of the Hamiltonian measured with IonQ device and simulated variance decreases as VQE progresses when close to convergence. Such trend is expected, as the variance of the Hamiltonian is zero when the output state is the ground state. Furthermore, we can see a similar trend between the variance of the Hamiltonian and HR distance measured with IonQ device(main Figure 2a and 2b) as VQE progresses.
	
	However, the range of the variance depends on the Hamiltonian. For the 11-qubit 1D-TFIM Hamiltonian, the variance of the Hamiltonian varies approximately between 6.5 and 12.5, but for the 6-qubit $J_{1}$-$J_{2}$ TFIM Hamiltonian, the variance of the Hamiltonian varies approximately between 2 and 6. However, the HR distance varies between 0 and 1, invariant of the Hamiltonian. This discrepancy shows how comparing the variance of the Hamiltonian becomes a case-by-case exercise to determine the value of variance indicative of optimal VQE convergence.
	
	\bibliographystyle{mcmahonlab.bst}
	\bibliography{references_main.bib}